\documentclass[aps,prb,twocolumn,10pt,superscriptaddress,amsmath,amsfonts,notitlepage,longbibliography]{revtex4-2}
\usepackage{amsmath,amsfonts,amssymb,dsfont,graphicx,bm,subcaption}
\usepackage[justification=Justified]{caption}
\graphicspath{{}}
\usepackage{color}
\usepackage{hyperref}
\usepackage{tikz}
\usepackage{pbox}
\usepackage{balance}

\usepackage{txfonts}

\usepackage[normalem]{ulem}

\def\be{\begin{equation}}
\def\ee{\end{equation}}
\def\bea{\begin{eqnarray}}
\def\eea{\end{eqnarray}}
\def\bpm{\begin{pmatrix}}
\def\epm{\end{pmatrix}}

\def\nn{\nonumber}

\newcommand{\p}{\partial}

\newcommand{\g}{\gamma}

\allowdisplaybreaks

\begin{document}
\title{Van-Hove singularities and competing instabilities in an altermagnetic metal}

\author{Peng Rao}
\affiliation{Physics Department, Technical University of Munich, TUM School of Natural Sciences, 85748 Garching, Germany}
\author{Johannes Knolle}
\affiliation{Physics Department, Technical University of Munich, TUM School of Natural Sciences, 85748 Garching, Germany}
\affiliation{Munich Center for Quantum Science and Technology (MCQST), Schellingstr. 4, 80799 München, Germany}
\affiliation{Blackett Laboratory, Imperial College London, London SW7 2AZ, United Kingdom}
\author{Laura Classen}
\affiliation{Physics Department, Technical University of Munich, TUM School of Natural Sciences, 85748 Garching, Germany}
\affiliation{Max-Planck-Institute for Solid State Research, 70569 Stuttgart, Germany}

\date{\today}
\begin{abstract}
Van-Hove (VH) singularities in the single-particle band spectrum are important for interaction-driven quantum phases. Whereas VH points are usually spin-degenerate, in newly proposed altermagnets VH singularities can become spin-dependent, due to momentum-dependent spin polarization of the Fermi surfaces arising from combined rotation and time-reversal symmetry. We consider two altermagnetic models ($d_{x^2-y^2}$- and $d_{xy}$-wave) on a square lattice with spin-polarized VH points, and study their stable fixed-point solutions indicating interaction-induced instabilities using parquet renormalization group. For both models, we find new stable fixed-point solutions of the renormalization group equations which are not connected to the solution in the spin-degenerate limit. This implies that on the square lattice, the system with VH singularities is unstable with respect to altermagnetic perturbations. The leading instability for the $d_{x^2-y^2}$-model is real transverse spin density wave. For the $d_{xy}$-wave model, it is found to be real transverse spin density wave at large altermagnetic splitting. At small altermagnetic splitting both imaginary charge density wave and real longitudinal spin density waves are dominant.
\end{abstract}

\maketitle

\section{Introduction}

A Van Hove (VH) scenario for competing orders, including superconductivity from repulsive interactions, was discussed for many correlated metals \cite{Furukawa1998,Hur2009,Chubukov2012,PhysRevB.104.195134,PhysRevX.8.041041,MARKIEWICZ19971179,Maitireview,hoVHs25}.
VH points correspond to saddle points in the energy dispersion in two-dimensional materials, at which two constant-energy contours in momentum space intersect. In 2D fermionic systems, when the Fermi energy $\varepsilon_F$ approaches a VH point, the density of states diverges logarithmically. This enhances the role of interactions and therefore, the system can exhibit many interaction-driven instabilities such as 
charge-density-waves (CDW), spin-density-waves (SDW), superconductivity (SC), or Pomeranchuk instabilities~\cite{MARKIEWICZ19971179,Maitireview,hoVHs25}. In particular, in systems with non-equivalent VH points and approximate nesting, e.g. on the square lattice at half filling~\cite{Furukawa1998,Hur2009}, and on the honeycomb lattice at quarter filling~\cite{Chubukov2012}, multiple divergent scattering channels can become coupled resulting in competing instabilities. 
However, in most cases the electron bands are taken to be spin-degenerate, since the relativistic spin-orbit coupling and the Zeeman splitting are usually weak compared to $\varepsilon_F$.

Recently, altermagnetism (AM) has been proposed to describe a class of materials in which collinear antiferromagnetic order with sublattice spin densities not related by translation or inversion but rather rotation symmetries \cite{Smejkal2022Sep}, indicating with strongly spin-split electron Fermi surfaces, the splitting being potentially comparable to the Fermi energy~\cite{Hayami2019,Yuan2020,ma2021multifunctional,Smejkal2022Dec,PhysRevB.99.184432,doi:10.1126/sciadv.aaz8809}.
Spin-polarized Fermi surfaces are also predicted in a number of theoretical scenarios such as the spin-channel Pomeranchuk instability~\cite{Pomeranchuk1958,Wu2007,Chubukov2018} and spin-nematic order in the Emery model~\cite{Fischer2011,li2024d}. 
In an altermagnet, despite the net magnitude of spin on each magnetic atom being equal, the local spin densities on sublattices of opposite spins are not related by translations or inversion due to, for example,  deformation by electrostatic interactions with the non-magnetic atoms or spontaneous orbital ordering~\cite{Valentin2024}. Therefore, unlike usual antiferromagnets, AM is invariant under combined point group and time-reversal symmetry~\cite{Smejkal2022Sep,Smejkal2022Dec}. It follows that the spin-splitting of Fermi surfaces in AM are non-zero, momentum-dependent, and can be large. Note that the AM mechanism is non-relativistic: spin components along the background N\'eel vector $\mathbf{n}$ are conserved and the system spin symmetry is reduced from SU$(2)$ to U$(1)$. The question therefore arises, whether spin-polarized VH singularities exist in AM, and what instabilities such VH points induce. In particular, if VH points are also present in the spin-degenerate limit, it is unclear if 
a finite AM splitting is a relevant perturbation for instabilities of the spin-degenerate limit.

\begin{figure}
    \centering
    \includegraphics[width=0.9\linewidth]{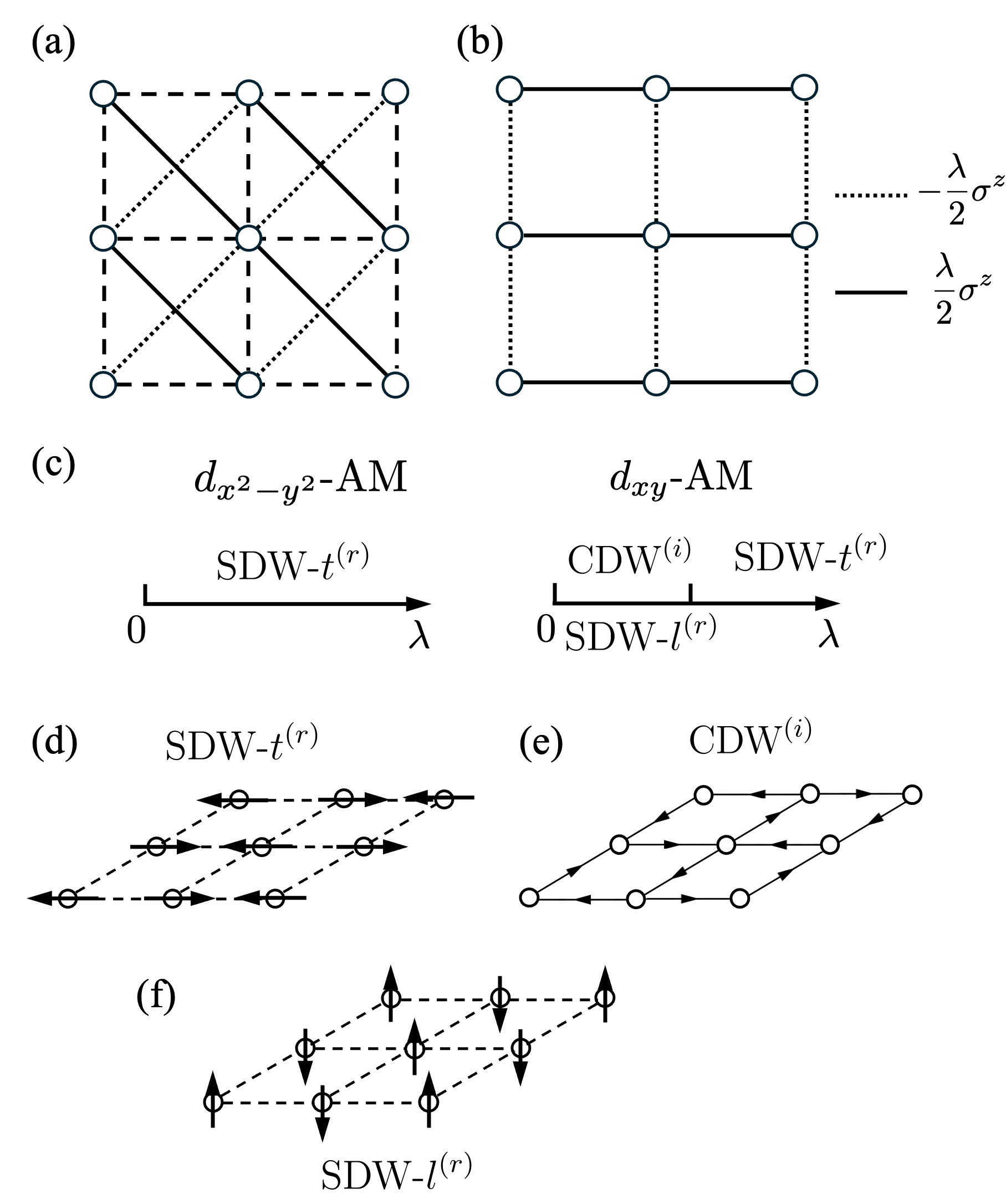}%
    \caption{(a)-(b): The effective models for Van Hove points of a square-lattice altermagnet [Eqs.~\eqref{eq:VHXxy},\eqref{eq:VHYxy},\eqref{eq:VHYx2-y2}, \eqref{eq:VHXx2-y2}] can have different microscopic realizations since they only rely on the symmetry-induced spin splitting of bands. A simple example is a single-site unit cell with altermagnetism-induced spin-dependent hoppings that realize (a) the $d_{xy}$-wave form factor \eqref{eq:form-factor-dxy}; (b) the $d_{x^2-y^2}$-wave form factor \eqref{eq:form-factor-d-wave} with altermagnetic vector $\mathbf{n}$ chosen along the z-direction. The system is invariant under $C_4\mathcal{T}$ transformation. (c) Schematic phase diagram of the two models as a function of the AM strength $\lambda$. The instabilities at $\lambda=0$ (not shown here) are not adiabatically connected to the phases shown. (d)-(e) schematic diagrams of the leading instabilities. (d) Real longitudinal spin-density-wave (l-SDW) corresponds to having AFM electronic spins collinear with $\mathbf{n}$. Besides translation, the l-SDW does not break any pure spin symmetry, but the $C_4\mathcal T$ (with on-site rotation center) of the AM phase. (e) Real transverse spin-density-wave (t-SDW) has AFM electronic spins in the plane perpendicular to $\mathbf{n}$. It breaks spin U(1), AM $C_4\mathcal T$, and translation symmetry. (f) The imaginary charge density wave gives different phases to the hopping of electrons and breaks translation symmetry~\cite{nayak2000density}.}
    \label{fig:schematic}
\end{figure}

In this study, we investigate the VH-induced instabilities of an altermagnetic metal on a square lattice, which is invariant under combined $4$-fold rotation and time-reversal $C_4\mathcal{T}$. 
Previous studies investigated if AM itself can be induced as an instability by spin-degenerate VH singularities \cite{yu2025altermagnetism,PhysRevB.111.L020401} or other electronic mechanisms \cite{Durrnagel2024Dec,PhysRevLett.132.263402,PhysRevB.110.144412,Liu2022Apr,li2024d,Valentin2024,Maier2023Sep}.
Here we assume that the system is already in the altermagnetic phase, and consider 
secondary instabilities. Altermagnets can have high transition temperatures due to their origin in the crystal anisotropy \cite{Smejkal2022Sep,Smejkal2022Dec,Fernandes2024Jan,Guo2023Mar} so that  an additional instability from inside the altermagnetic state becomes a  realistic possibility at lower temperature. Recently, such a secondary instability in the form of a spin density wave was reported in experiments of the room-temperature altermagnet KV$_2$Se$_2$O \cite{jiang2025metallic}. Unusual forms of superconductivity in an AM parent state were also studied either by assuming an attractive interaction \cite{sim2024PDW,PhysRevB.108.184505,Zhang2024finite,PhysRevB.110.L060508,PhysRevB.110.024503} or deriving it microscopically~\cite{Brekke2023Dec,bose2024tJ}.
We consider two $C_4\mathcal{T}$ invariant tight-binding models with $d_{x^2-y^2}$- or $d_{xy}$-wave AM and repulsive electronic interactions; see Fig.~\ref{fig:schematic}(a)-(b). We 
find that both models contain spin-
dependent VH points at high-symmetry points $X = (\pi,0)$ and $Y=(0,\pi)$ in the Brillouin zone (BZ). In the $d_{x^2-y^2}$-wave model, which is hypothesized to be realized in AM candidate materials such as or the V$_2$X$_2$O (X = Te, Se) family~\cite{Libor2022,ma2021multifunctional,jiang2025metallic,parthenios2025}, each VH point has only a Fermi surface of one spin-component and the corresponding $\varepsilon_F$ is shifted away from zero. In the $d_{xy}$-wave model suggested to be realised in CoS$_2$ \cite{parthenios2025}, both spin components form VH points at $X$ and $Y$ at $\varepsilon_F=0$. In both models, at vanishing AM strength $\lambda$, 
spin-up and -down Fermi surfaces coincide and the system tends to the usual SU$(2)$-symmetric limit, 
including spin-degenerate VH-points. 

To study the VH-induced interaction effects, we show that in both models, the leading divergent one-loop diagrams correspond to scattering near each VH point with vanishing total momentum (the particle-particle channel) and scattering across $X$ and $Y$ with momentum transfer close to $\mathbf{Q}= (\pi,\pi)$ (the particle-hole channel). We then use a patch model which considers only electrons near the VH points and identify all independent coupling constants for both models. We solve their parquet renormalization group (pRG) equations which automatically sum 
the leading diagrams to logarithmic accuracy. For both the $d_{xy}$-wave and the $d_{x^2-y^2}$-wave models, we find two stable fixed-point solutions. They correspond to divergences in the coupling constants with predominantly anti-parallel incoming electron spin, and in couplings with parallel incoming spins. These solutions do not become the SU$(2)$-symmetric $d$-wave singlet superconductivity solution~\cite{Furukawa1998} at vanishing $\lambda$, the latter appearing as an unstable fixed-point solution to the pRG equations. This suggests that the system is unstable with respect to SU$(2)$-breaking AM perturbations.    

The one-loop pRG approximation allows us to identify the leading instabilities in the weak coupling limit. Since a non-zero order parameter for continuous symmetry breaking in 2D is forbidden at finite temperatures, by instabilities we mean the divergence of the corresponding susceptibilities under pRG. At zero temperature, these are true instabilities signaling a phase transition towards a phase of spontaneously broken symmetry. The exceptions are layered 2D materials where the broken-symmetry phase is stabilized by inter-layer coupling. We show that real SDW with spin transverse to the AM direction $\mathbf{n}$ is dominant for $d_{x^2-y^2}$-wave AM, whereas spin-singlet superconductivity is completely suppressed. For $d_{xy}$-wave AM, real transverse SDW becomes the dominant instability at large $\lambda$. As $\lambda$ decreases, real SDW parallel along $\mathbf{n}$ and imaginary CDW have the strongest divergence in susceptibility; we expect the `degeneracy' of the two orders to be lifted by including subleading diagrams. The results are summarized in a schematical phase diagram in Fig.~\ref{fig:schematic}(c). We also show the instabilities schematically in Fig.~\ref{fig:schematic}(d)-(f). This ordering occurs on top of the altermagnetic background. The precise physical picture of the SDW instabilities depend on the sublattice structure, and cannot be deduced from the effective  Van Hove model we use based on symmetry alone. For example, longitudinal SDW on the Lieb lattice results in unequal magnitude of the spin up and down on sublattices~\cite{jiang2025metallic}. Similarly the real transverse SDW on such a lattice causes the `tilting' of the N\'eel order away from the collinear configuration. However, generically the VH points should be gapped by the density wave instabilities. We emphasize that our analysis makes use of new pRG solutions due to AM, and the instabilities are not connected to the SU$(2)$-limit, in which $d$-wave singlet superconductivity is the dominant instability~\cite{Furukawa1998}. Our results agree qualitatively with Ref.~\onlinecite{parthenios2025} which considers the same models using the functional renormalization group (fRG) method. In this paper we only consider singlet superconductivity. Triplet superconductivity requires a momentum-dependent interaction vertex, and cannot be studied within the current patch model; see Appendix~\ref{sec:triplet}. See also Ref.~\onlinecite{ojajarvi2024pairing} for the case of a single VH point. However, we argue that one of the PRG solutions leaves the possibility open that a triplet pairing instability can occur.

The rest of the paper is organized as follows. In Sec.~\ref{sec:bandstructure} we introduce the two AM models and compute the most divergent one-loop diagrams. In Sec.~\ref{sec:patch-RG} we identify the independent coupling constants in both models and derive the coupled RG equations. We present their numerical solutions 
corresponding to the two aforementioned stable fixed points. Lastly, we consider the vertex RG equations for possible instabilities in Sec.~\ref{sec:susceptibilities} 
and compute the susceptibilities 
for the stable fixed points to determine the corresponding types of instability.

\section{\label{sec:bandstructure}Model \& bare susceptibilities}
A minimal model for the bands of an AM metal on a square lattice is given by the tight binding Hamiltonian in quasi-momentum space \cite{Smejkal2022Dec}
\begin{equation}\label{eq:TB-Hamiltonian}
    H_0(\mathbf{p}) = -2t \left(\cos p_x + \cos p_y \right) - \mu +   \lambda f(\mathbf{p}) \sigma^z,
\end{equation}
where $\mu$ is the chemical potential, $\lambda$ is the AM strength and $\sigma^z$ is the third Pauli matrix for spin. Throughout this paper we take the lattice constant to be unity. The last term in \eqref{eq:TB-Hamiltonian} is due to the altermagnetic background, which we take 
to be along the $z$-direction), and splits the Fermi surface into spin up and down components. Note that the spin components along $\mathbf{e}_z$ are still conserved. Due to the $C_4\mathcal{T}$ symmetry, the function $f(\mathbf{p})$ is odd under a $C_4$ rotation.

In the following, we consider two possible choices of $f(\mathbf{p})$ form factors. As will be shown below, both models result in VH singularities in the band-structure at high symmetry points $X = (\pi,0)$ and $Y=(0,\pi)$ for $\lambda< 2t$.

\subsection{\texorpdfstring{$d_{xy}$}{TEXT}-wave model-wave model}

\begin{figure}
    \centering
    \includegraphics[width=\linewidth]{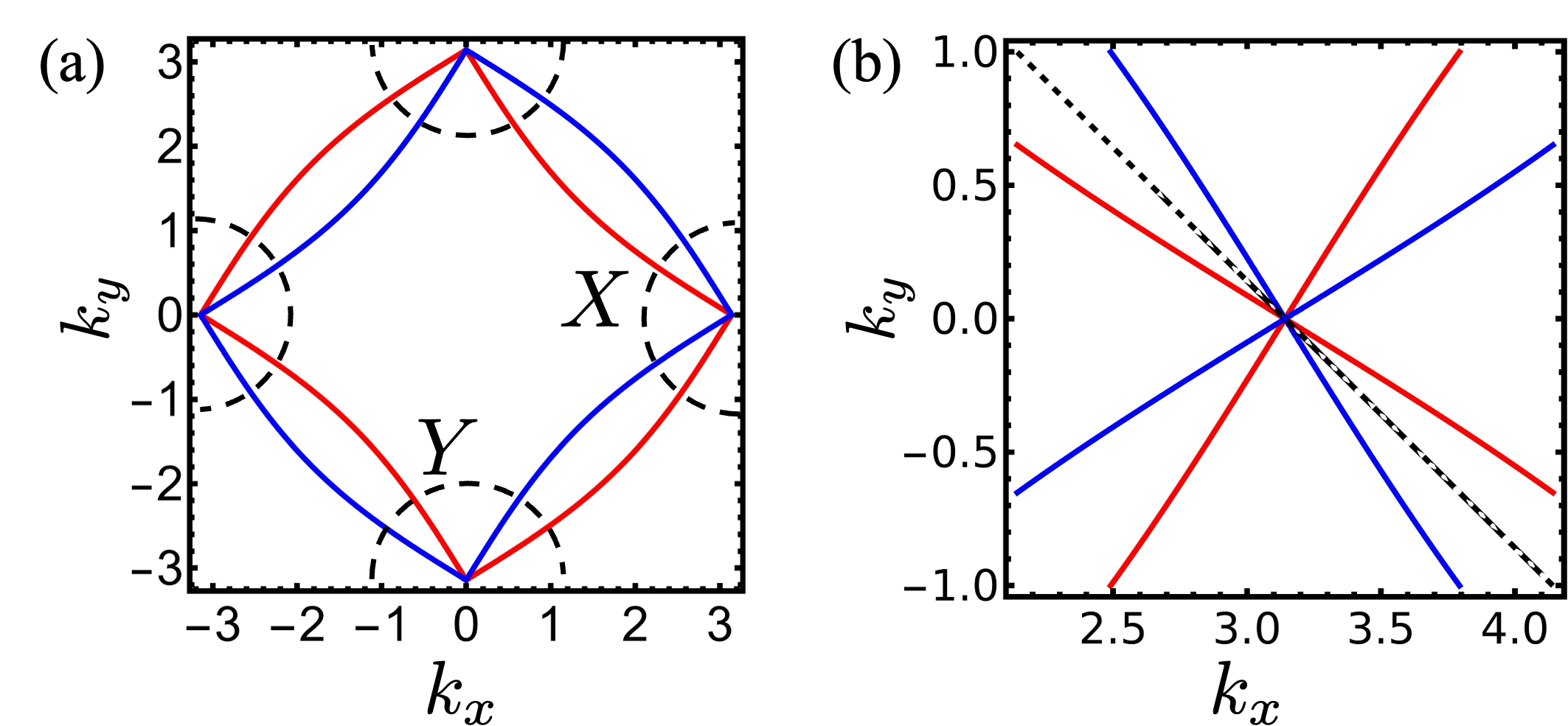}%
    \caption{Fermi surface of the $d_{xy}$-wave model at $\mu=0$ and $\lambda = 0.5 t$: (a) on the entire Brillouin zone; (b) near the patch $X$. Red and blue represents spin up and down respectively. The Van Hove points are located at the centers of patch $X,Y$, e.g. the origin in (b). The areas bounded by dashed lines of size $\Lambda$} are the patches.
    \label{fig:Fermi-surface-dxy}
\end{figure}

We first consider the following $C_4$-odd form-factor:
\begin{equation}\label{eq:form-factor-dxy}
    f(\mathbf{p}) = \sin p_x \sin p_y.
\end{equation}
Similar to the spin-degenerate case, the VH singularities occur at $\mu = 0$ as shown in Fig.~\ref{fig:Fermi-surface-dxy}(a). Near each VH point, altermagnetic splitting rotates the Fermi-surface of opposite spin components by opposite angles; see Fig.~\ref{fig:Fermi-surface-dxy}(b). Near $X=(\pi,0)$ and $Y=(0,\pi)$ we can approximate the dispersion in a patch of size $\ll 1$ via
\begin{align}
    \varepsilon_{X,\sigma}(\mathbf{p})&=  -t(p_x^2-p_y^2) - \lambda \sigma p_xp_y, \label{eq:VHXxy}\\
    \varepsilon_{Y,\sigma}(\mathbf{p})&=  t(p_x^2-p_y^2) - \lambda \sigma p_xp_y \label{eq:VHYxy}
\end{align}
with $\sigma=\pm$.

We observe that Fermi surfaces of opposite spins have perfect nesting across VH points
\begin{equation}\label{eq:nesting-dxy-1}
    \varepsilon_{X,\sigma}(\mathbf{p}) = - \varepsilon_{-\sigma,Y}(\mathbf{p}) 
\end{equation}
whereas for parallel spins the nesting is broken by the AM term
\begin{equation}\label{eq:nesting-dxy-2}
    \varepsilon_{X,+}(\mathbf{p}) = - \varepsilon_{Y,+}(\mathbf{p})-2\lambda p_xp_y ,
\end{equation}

To analyse the propensity of the AM metal towards potential instabilities, we calculate the bare particle-hole and particle-particle susceptibilities. These will also enter in the interacting case as one-loop corrections for the dressed interaction. The bare susceptibilities are given by one-loop diagrams and we find that they are divergent for specific momentum transfers.  
The leading divergent one-loop diagrams are the particle-particle diagram $\Pi_{\text{pp}}$ near each VH point and the particle-hole diagram $\Pi_{\text{ph}}$ across VH points with momentum transfer $\mathbf{Q}=(\pi,\pi)$, for both parallel (p) and anti-parallel (ap) incoming electron spins. We find that $\Pi^{\text{(p)}}_{\text{pp}}$ and $\Pi^{\text{(ap)}}_{\text{ph}}$ contain double logarithmic divergences
\begin{subequations}\label{eq:one-loop-dxy}
\begin{align}
    &\Pi^{\text{(p)}}_{\text{pp}}=  -\frac{1}{8\pi^2 \sqrt{t^2+(\lambda/2)^2}} \log^2 \left( \frac{\Lambda}{\max \{\omega,T \}}\right), \label{eq:one-loop-dxy-BCS} \\
    &\Pi^{\text{(ap)}}_{\text{ph}} = -\Pi^{\text{(p)}}_{\text{pp}},\label{eq:one-loop-dxy-nesting}
\end{align}
\end{subequations}
where $\omega$ is the external frequency and $T$ the temperature. $\Lambda \sim t$ is the UV cut-off. The double logarithmic divergence in $\Pi^{\text{(ap)}}_{\text{ph}}$ is due to the aforementioned perfect nesting of FS with anti-parallel spins, Eq.~\eqref{eq:nesting-dxy-1}, and the logarithmic density of states. The analogous divergence in $\Pi^{\text{(ap)}}_{\text{pp}}$ arises due to the combination of the Cooper logarithm and the density of states. For $\Pi^{\text{(ap)}}_{\text{pp}}$ and $\Pi^{\text{(p)}}_{\text{ph}}$ diagrams, one logarithm is cut by the distortion of the Fermi surface from the AM term in Eq.~\eqref{eq:nesting-dxy-2}
\begin{subequations}\label{eq:one-loop-dxy-1}
\begin{align}
    &\Pi^{\text{(ap)}}_{\text{pp}}=  -  \frac{1}{4\pi^2t} \log \left(\frac{\Lambda}{\max \{\omega,T \}}\right) \log \left(\frac{\max \{\omega,\lambda\}}{t}\right),\label{eq:one-loop-dxy-1-BCS} \\
    &\Pi^{\text{(p)}}_{\text{ph}} = -\Pi^{\text{(ap)}}_{\text{pp}}.\label{eq:one-loop-dxy-1-nesting}
\end{align}
\end{subequations}
The details of deriving Eqs.~\eqref{eq:one-loop-dxy} and \eqref{eq:one-loop-dxy-1} are given in Appendix~\ref{sec:one-loop}.

\subsection{\texorpdfstring{$d_{x^2-y^2}$}{TEXT}-wave model}

\begin{figure}
    \centering
    \includegraphics[width=\linewidth]{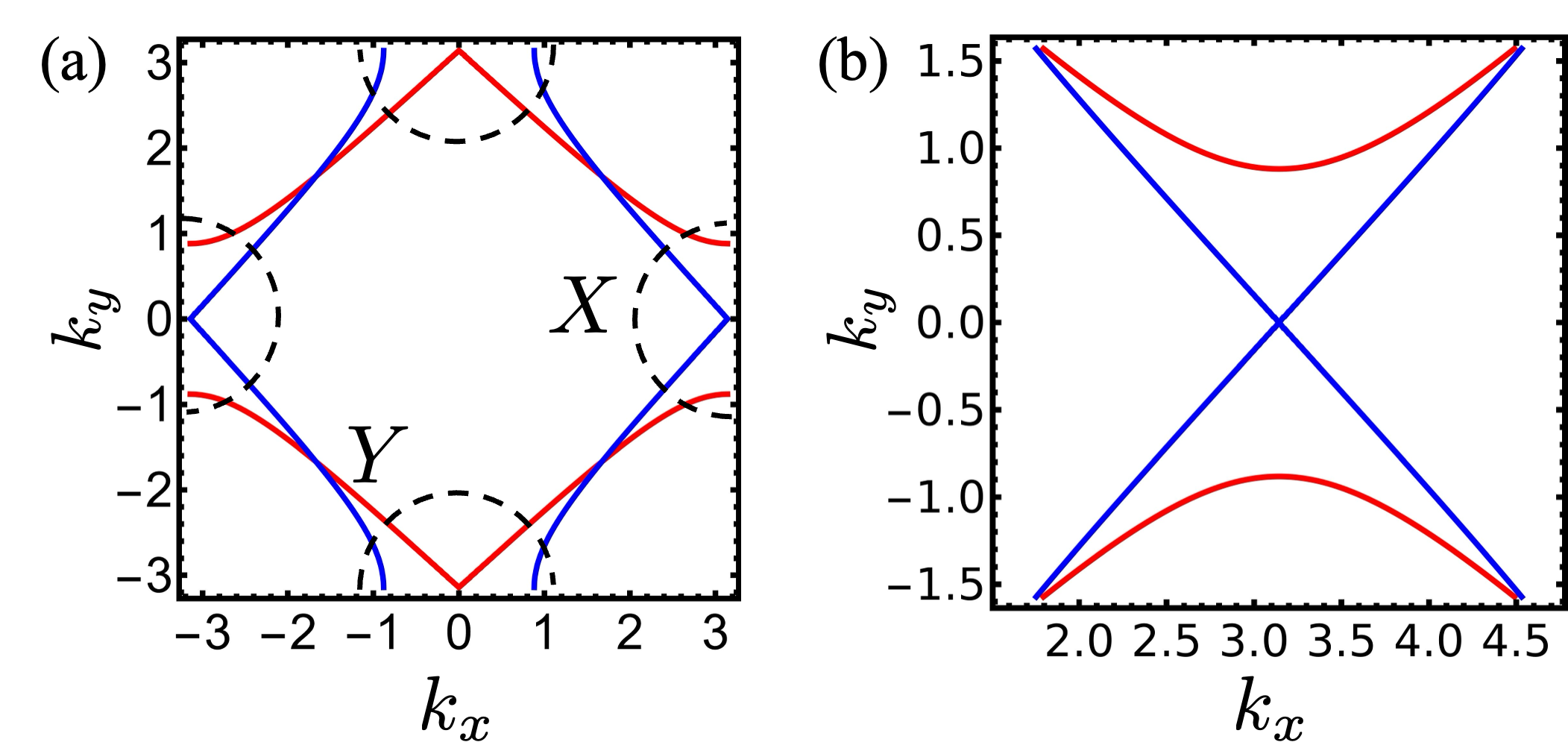}%
    \caption{Fermi surface of the $d_{x^2-y^2}$-wave model at $\lambda = 0.1 t$ and  $\mu=2\lambda$: (a) on the entire BZ; (b) near the patch $X$. Red and blue represents spin up and down respectively. Only the spin up (down) VH point exists at patch $X$ ($Y$).  The areas bounded by dashed lines of size $\Lambda$ are the patches.}
    \label{fig:Fermi-surface-d-wave}
\end{figure}

Aside from Eq.~\eqref{eq:form-factor-d-wave}, there exists another $C_4$-odd form factor which is predicted in candidate AM materials~\cite{Smejkal2022Sep}:
\begin{equation}\label{eq:form-factor-d-wave}
    f(\mathbf{p}) = \cos p_x - \cos p_y.
\end{equation}
In Sec.~\ref{sec:exp-scenario} we show that Eq.~\eqref{eq:form-factor-d-wave} can be realized in the Emery model as originally proposed by \cite{Fischer2011}. Contrary to the $d_{x^2-y^2}$-wave AM, the altermagnetic term shifts the VH singularities from $\mu=0$ to $\mu = \pm 2\lambda$. The bandstructure for $\mu = 2\lambda$ is shown in Fig.~\ref{fig:Fermi-surface-d-wave}(a). We see that near each patch only one spin-component has VH singularities. This is shown for the spin-down FS near patch $X$ in Fig.~\ref{fig:Fermi-surface-d-wave}(b). For concreteness we shall consider $\mu = 2\lambda$ in what follows.

Near patch $Y$ the spin-up Fermi surface contains a VH-point:
\begin{equation} \label{eq:VHYx2-y2}
    \varepsilon_{Y,+}(\mathbf{p})=  \left(t-\frac{\lambda}{2}\right)p_x^2- \left(t+\frac{\lambda}{2}\right)p_y^2,
\end{equation}
while for spin-down electrons the Fermi surface is separated from the VH-point due to the AM splitting:
\begin{equation}
    \varepsilon_{Y,-}(\mathbf{p})=  \left(t+\frac{\lambda}{2}\right)p_x^2- \left(t-\frac{\lambda}{2}\right)p_y^2-2\mu.
\end{equation}
Fermi-surfaces at patch $X$ can be obtained by the $C_4\mathcal{T}$ transformation
\begin{equation} \label{eq:VHXx2-y2}
    \varepsilon_{X,\sigma}(\mathbf{p})=\varepsilon_{Y,-\sigma}(C_4\mathbf{p}).
\end{equation} 
Here the perfect nesting of Fermi surfaces across VH points is broken by the AM term 
\begin{equation}\label{eq:nesting-d-wave}
    \varepsilon_{X,-}(\mathbf{p}) = - \varepsilon_{Y,+}(\mathbf{p})- \lambda p^2.
\end{equation}

We now calculate again the bare particle-particle and particle-hole susceptibilities. To this end, we consider the one-loop diagrams in the Matsubara representation. The most divergent one-loop diagrams are the particle-particle diagram $\Pi_{\text{pp}}$ for spin-down at $X$ and spin-up at $Y$ due to the VH points, and the particle-hole diagram $\Pi_{\text{ph}}$ with spin-down at $X$ and spin-up at $Y$ across VH points with momentum transfer $\mathbf{Q}=(\pi,\pi)$. In the $\lambda \ll t$ limit they are given by
\begin{subequations}\label{eq:one-loop-d-wave}
\begin{align}
    &\Pi_{\text{pp}}=  -   \frac{1}{8\pi^2t} \log^2 \left(\frac{\Lambda}{\max \{\omega,T \}}\right) ;\label{eq:one-loop-d-wave-BCS}\\
    &\Pi_{\text{ph}} = \frac{1}{4\pi^2t} \log \left(\frac{\Lambda}{\max \{\omega,T \}}\right) \log \left(\frac{\max \{\omega,\lambda\}}{t}\right).\label{eq:one-loop-d-wave-nesting}
\end{align}
\end{subequations}
The particle-particle diagram $\Pi_{\text{pp}}$ has double-logarithmic divergences, whereas the particle-hole diagram $\Pi_{\text{ph}}$ has a single logarithmic divergence due to the imperfect nesting by AM; see Eq.~\eqref{eq:nesting-d-wave}. We are unable to derive analytical expressions for other one-loop diagrams. But in these diagrams both logs are cut by the AM splitting $\lambda$ which can be estimated as:
\begin{equation}\label{eq:one-loop-d-wave-subleading}
  \Pi  \sim \text{const.}\log^2\left(\frac{\max \{\omega,\lambda\}}{t}\right);
\end{equation}
their coefficients can be different however for different diagrams. The details for deriving Eqs.~\eqref{eq:one-loop-d-wave} and \eqref{eq:one-loop-d-wave-subleading} are given in Appendix~\ref{sec:one-loop}.

\begin{figure}
    \centering
    \includegraphics[width=0.9\linewidth]{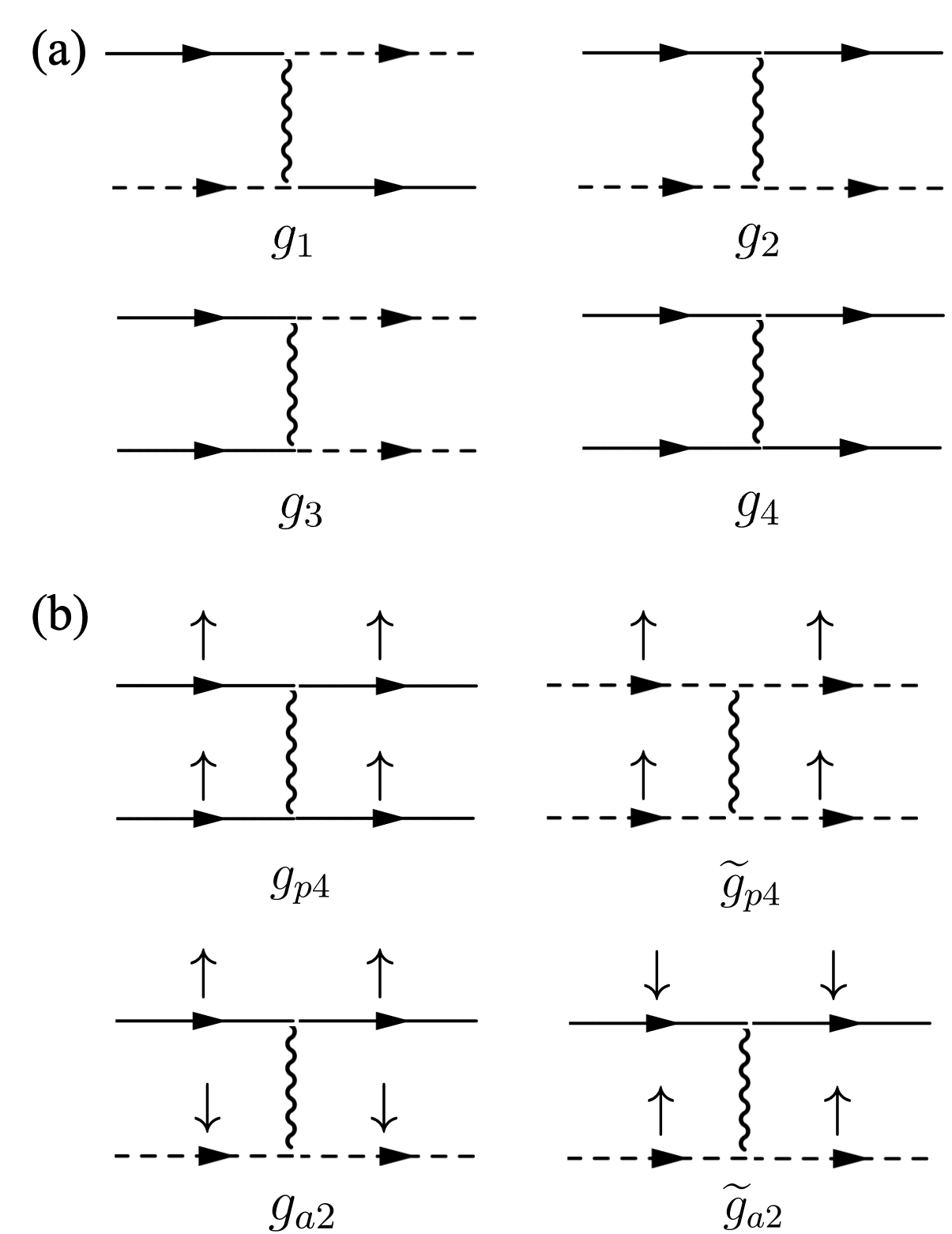}%
    \caption{The scattering channels for coupling constants in systems with two VH points. (a) couplings without external spin indices. Solid lines correspond to electrons at patch $X$ and dashed lines to electrons at patch $Y$. Putting in parallel or anti-parallel electron spins gives the coupling constants in Eq.~\eqref{eq:coupling-constants}. In (b) we show the $g_{p4}, \widetilde{g}_{p4}$ and $ g_{a2}, \widetilde{g}_{a2}$ couplings which are in the same channels but are not related by $C_4 \mathcal{T}$ symmetry.}
    \label{fig:coupling-constants}
\end{figure}

\section{Symmetry-allowed interactions and  parquet 
RG}\label{sec:patch-RG}
The singularities of bare susceptibilities, which we found in the previous section indicate that the AM metal at VH filling is unstable upon the addition of interactions and that there is a competition between different ordering tendencies because particle-particle and particle-hole susceptibilities are of the same size.
In what follows we determine all independent, symmetry-allowed couplings at and between VH points for the two cases of $d_{xy}$ and $d_{x^2-y^2}$ AM splitting. We then use parquet RG~\cite{Furukawa1998,Chubukov2012} to study the leading instabilities. This is equivalent to summing over all aforementioned leading diagrams for the corresponding scattering vertices, i.e., it is exact to order $\log^2$.

\subsection{Independent coupling constants}

The presence of distinct spin up and down Fermi surfaces near each VH point necessitates considering, separately, scattering channels for incoming electrons with parallel or anti-parallel spins. We make use of the remaining U$(1)$ spin symmetry, which allows us to express all scattering processes in terms of interaction vertices which conserve the spin components along each electron line. Together with $C_4\mathcal{T}$ invariance this leads to the following $10$ independent coupling constants
\begin{equation}\label{eq:coupling-constants}
\begin{split}
    \text{parallel spin}:& \ g_{p1}, g_{p2},g_{p3},g_{p4}, \widetilde{g}_{p4}; \\
    \text{anti-parallel spin}:& \ g_{a1}, g_{a2}, \widetilde{g}_{a2}, g_{a3},g_{a4}. 
\end{split}
\end{equation}
They correspond to the following interaction vertices for parallel-spin scattering
\begin{equation}
\begin{split}
   H_{\mathrm{int},p} &= \frac{1}{2}\bigg[\sum_{\sigma=\pm}\sum_{\alpha\ne \beta} \bigg(g_{p1}\psi_{\alpha,\sigma}^{\dagger}\psi_{\beta,\sigma}^{\dagger}\psi_{\alpha,\sigma} \psi_{,\sigma} + g_{p2}\psi_{\alpha,\sigma}^{\dagger}\psi_{\beta,\sigma}^{\dagger} \psi_{\beta,\sigma} \psi_{\alpha,\sigma} \\
    &+ g_{p3}\psi_{\alpha,\sigma}^{\dagger}\psi_{\alpha,\sigma}^{\dagger} \psi_{\beta,\sigma} \psi_{\beta,\sigma} \bigg) \\
    &+ g_{p4}\left( \psi_{X,+}^{\dagger}\psi_{X,+}^{\dagger} \psi_{X,+} \psi_{X,+} +\psi_{Y,-}^{\dagger}\psi_{Y,-}^{\dagger} \psi_{Y,-} \psi_{Y,-}\right) \\
    &+ \widetilde{g}_{p4}\left( \psi_{X,-}^{\dagger}\psi_{X,-}^{\dagger} \psi_{X,-} \psi_{X,-} +\psi_{Y,+}^{\dagger}\psi_{Y,+}^{\dagger} \psi_{Y,+} \psi_{Y,+}\right)\bigg].
\end{split}
\end{equation}
and 
for anti-parallel spin scattering
\begin{equation}
\begin{split}
    H_{\mathrm{int},ap} =&\frac{1}{2}\bigg[\sum_{\sigma\ne \sigma'} \sum_{\alpha\ne \beta} \bigg( g_{a1}\psi_{\alpha,\sigma}^{\dagger}\psi_{\beta,\sigma'}^{\dagger}\psi_{\alpha,\sigma'} \psi_{,\sigma}\\
    &+ g_{a3}\psi_{\alpha,\sigma}^{\dagger}\psi_{\alpha,\sigma'}^{\dagger} \psi_{\beta,\sigma'} \psi_{\beta,\sigma} + g_{a4}\psi_{\alpha,\sigma}^{\dagger}\psi_{\alpha,\sigma'}^{\dagger} \psi_{\alpha,\sigma} \psi_{\alpha,\sigma'}\bigg)  \bigg]\\
    & + g_{a2}\psi_{X,+}^{\dagger}\psi_{Y,-}^{\dagger} \psi_{Y,-} \psi_{X,+} +  \widetilde{g}_{a2}\psi_{Y,+}^{\dagger}\psi_{X,-}^{\dagger} \psi_{X,-} \psi_{Y,+}.
\end{split}
\end{equation}
The spin indices are $\sigma,\sigma'$ and $\alpha, \beta$ are patch indices. Electron momenta are not written out explicitly. 
The scattering channels are shown in Fig.~\ref{fig:coupling-constants} where electrons at patch $X$ are shown as solid lines and at patch $Y$ as dashed lines. Other types of interactions such as $(\psi^\dagger \sigma_\pm \psi)^2$ can be represented in terms of the ones above in terms of the completeness relation for Pauli matrices. In the spin-degenerate SU$(2)$ limit, we have $\widetilde{g}_{p4}=g_{p4}$, $\widetilde{g}_{a2}=g_{a2}$ and $g_{pi}=g_{ai}=g_i$.

\subsection{PRG equations}

\begin{figure*}
    \centering
    \includegraphics[width=\linewidth]{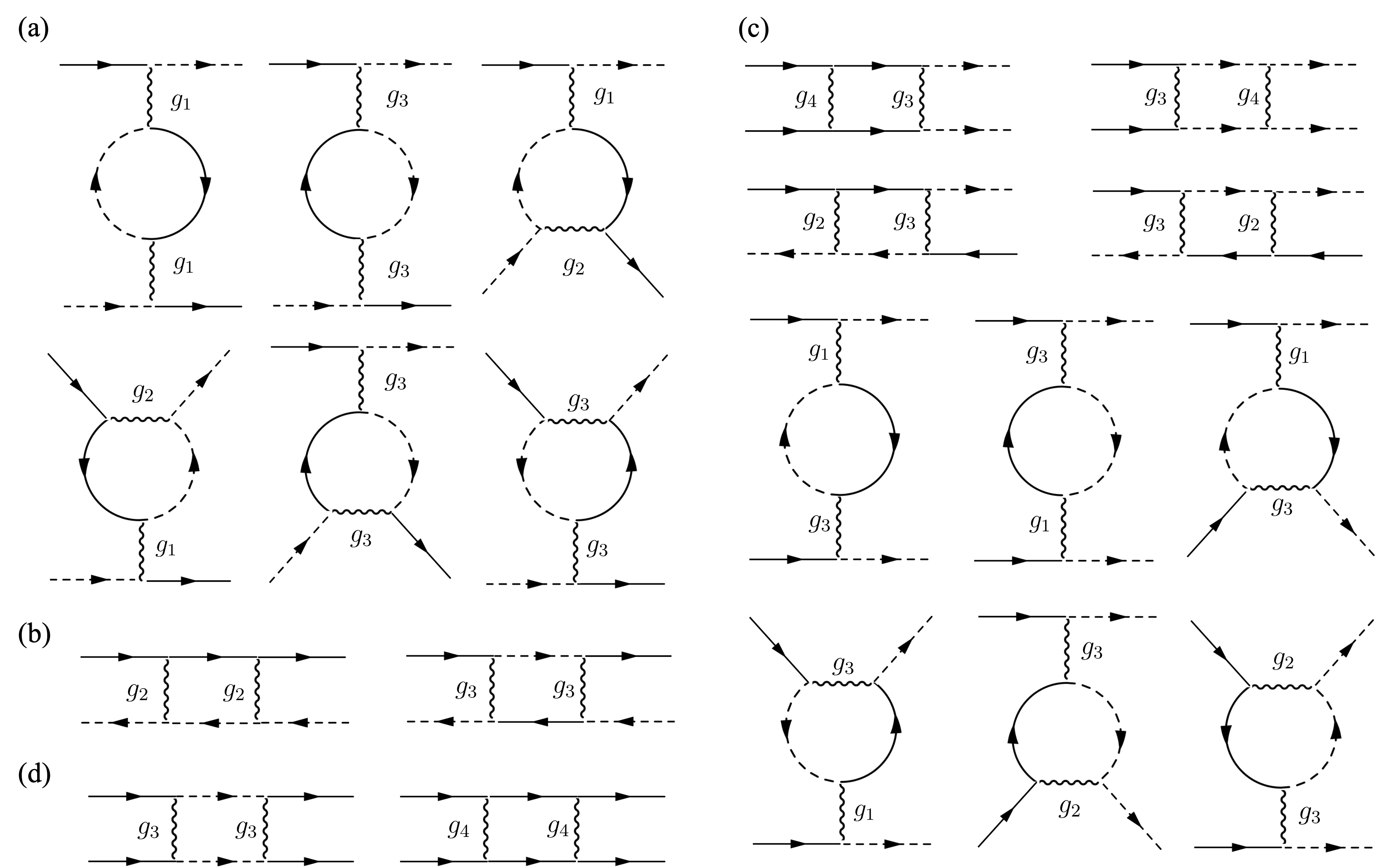}%
    \caption{One-loop diagrams for the coupling constants Eq.~\eqref{eq:coupling-constants}. (a)-(d) correspond to $g_{i1}, g_{i2},g_{i3}, g_{i4}$ where $i = p,a$; $\widetilde{g}_{a2}$ and $\widetilde{g}_{p4}$ have the same diagrams as $g_{a2}$ and $g_{p4}$. Solid lines correspond to patch $X$ while dashed lines to patch $Y$. The spin indices are not shown explicitly. The pRG equations \eqref{eq:RG-dxy} and \eqref{eq:RG-d-wave-full} are obtained by setting the incoming electron spin indices to be parallel and anti-parallel respectively and using the one-loop bare susceptibilities in Sec.~\ref{sec:bandstructure}.}
    \label{fig:RG-diagrams-dxy}
\end{figure*}

The couplings become dressed by multiple scattering events in particle-particle and particle-hole channels. Since the correction in the different channels are of the same order, we use pRG to account for them on equal footing. This allows us to systematically sum the infinite set of pure and mixed diagrams that contain the leading logarithmic behavior on every loop order. We derive the pRG equations for the two models below. The relevant Feynman diagrams for the coupling constants are shown in Fig.~\ref{fig:RG-diagrams-dxy}. 

\subsubsection{\texorpdfstring{$d_{xy}$}{TEXT}-wave model}

Taking into account the one-loop diagrams for the $d_{xy}$-wave AM leads to the following pRG equations  
\begin{subequations}\label{eq:RG-dxy}
\begin{align}
    &g_{p1}'= d [-(g_{p1})^2-(g_{a1})^2+ (g_{p3})^2 - (g_{a3})^2+ 2 g_{p1}g_{p2}]\\
    &g_{p2}'= d [(g_{p2})^2+(g_{p3})^2] \\
    &g_{p3}'= - g_{p3}(g_{p4}+\widetilde{g}_{p4}) + 2d(2g_{p2}g_{p3} -  g_{a1}g_{a3})\\
    &g_{p4}'= - (g_{p3})^2 - (g_{p4})^2\\
    &\widetilde{g}_{p4}'= - (g_{p3})^2 - (\widetilde{g}_{p4})^2\\
    &g_{a1}'= -2d (g_{p1}g_{a1}-g_{a1}g_{p2}) \\
    &g_{a2}'= (g_{a2})^2 + (g_{a3})^2 \\
    &\widetilde{g}_{a2}'= (\widetilde{g}_{a2})^2 + (g_{a3})^2 \\
    &g_{a3}'=- 2 d  ( g_{a3} g_{a4}+ g_{p1} g_{a3}-g_{p2} g_{a3}) + (g_{a2}+\widetilde{g}_{a2}) g_{a3}\\
    &g_{a4}'=-  d [ (g_{a3})^2+ (g_{a4})^2]
\end{align}
\end{subequations}
We have introduced the double logarithmic variable $y=\Pi^\text{(ap)}_\text{ph}$ and the $\beta$ functions $\beta_i=g_i'=\partial g_i/\partial y$. The function $d(y) = \mathrm{d}\Pi^\text{(p)}_\text{ph}/\mathrm{d}y$ describes the reduction of the double-logarithm to a large single logarithm due to AM in the corresponding diagrams. It 
satisfies $0<d(y)\le1$. For our numerical calculations, we set $d(y)=1/\sqrt{1+y}$~\cite{Chubukov2012} since at small $y$, $\Pi^\text{(p)}_\text{ph} =y$ and at large $y$, $\Pi^\text{(p)}_\text{ph} \sim \sqrt{y}$; see Eq.~\eqref{eq:one-loop-dxy}.

\begin{figure}
    \centering
    \includegraphics[width=\linewidth]{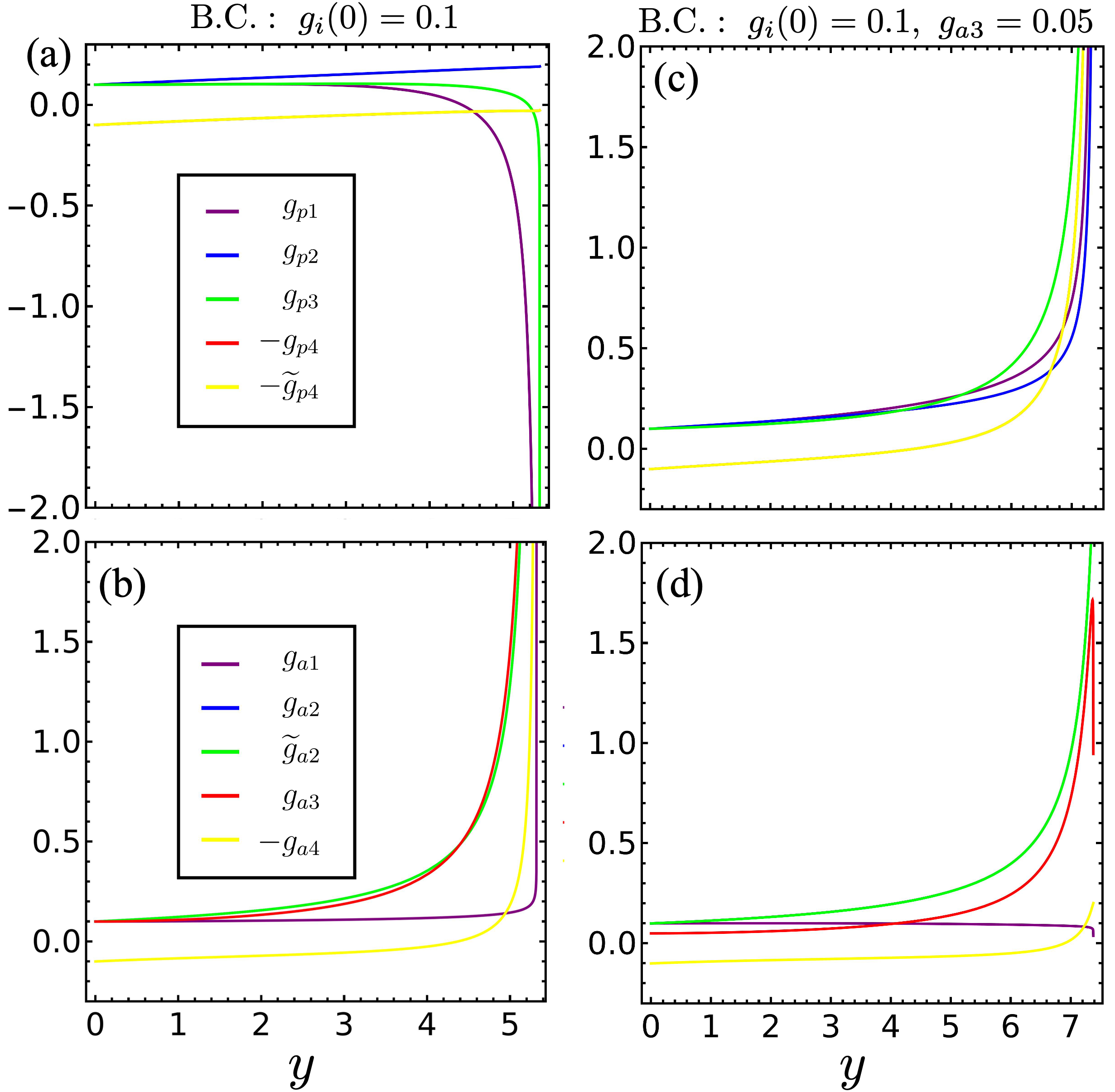}%
    \caption{Numerical solutions to the RG solutions \eqref{eq:RG-dxy} for $d_{xy}$-wave AM. Upper and lower rows show the results for $g_{pi}(y)$ in (a), (c); and $g_{ai}(y)$ in (b), (d) respectively. In (a) and (b) the initial conditions are $g_{p}(0) =g_{a}(0) = 0.1$; (c), (d) have the same initial conditions apart from $g_{a3}=0.05$.}
    \label{fig:RG-dxy}
\end{figure}

We present the numerical solutions to the RG equations in Fig.~\ref{fig:RG-dxy} By changing the 
initial conditions, 
we found two limiting behaviors corresponding to divergences in anti-parallel spin couplings and $g_{p1}$ [Fig.~\ref{fig:RG-dxy}(a)-(b)], or divergences in parallel spin couplings only (Fig.~\ref{fig:RG-dxy}(c)-(d)). We obtain the anti-parallel solution for all inital $g_i(0)=0.1$, and the parallel one for $g_{a3}(0) = 0.1, 0.05$ while fixing other $g_{i}(0) = 0.1$. This is expected since $g_{a3}$ is doubly-logarithmically divergent and its coupling to $g_{a4}$ governs the latter's divergence. 

\begin{figure}
    \centering
    \includegraphics[width=0.8\linewidth]{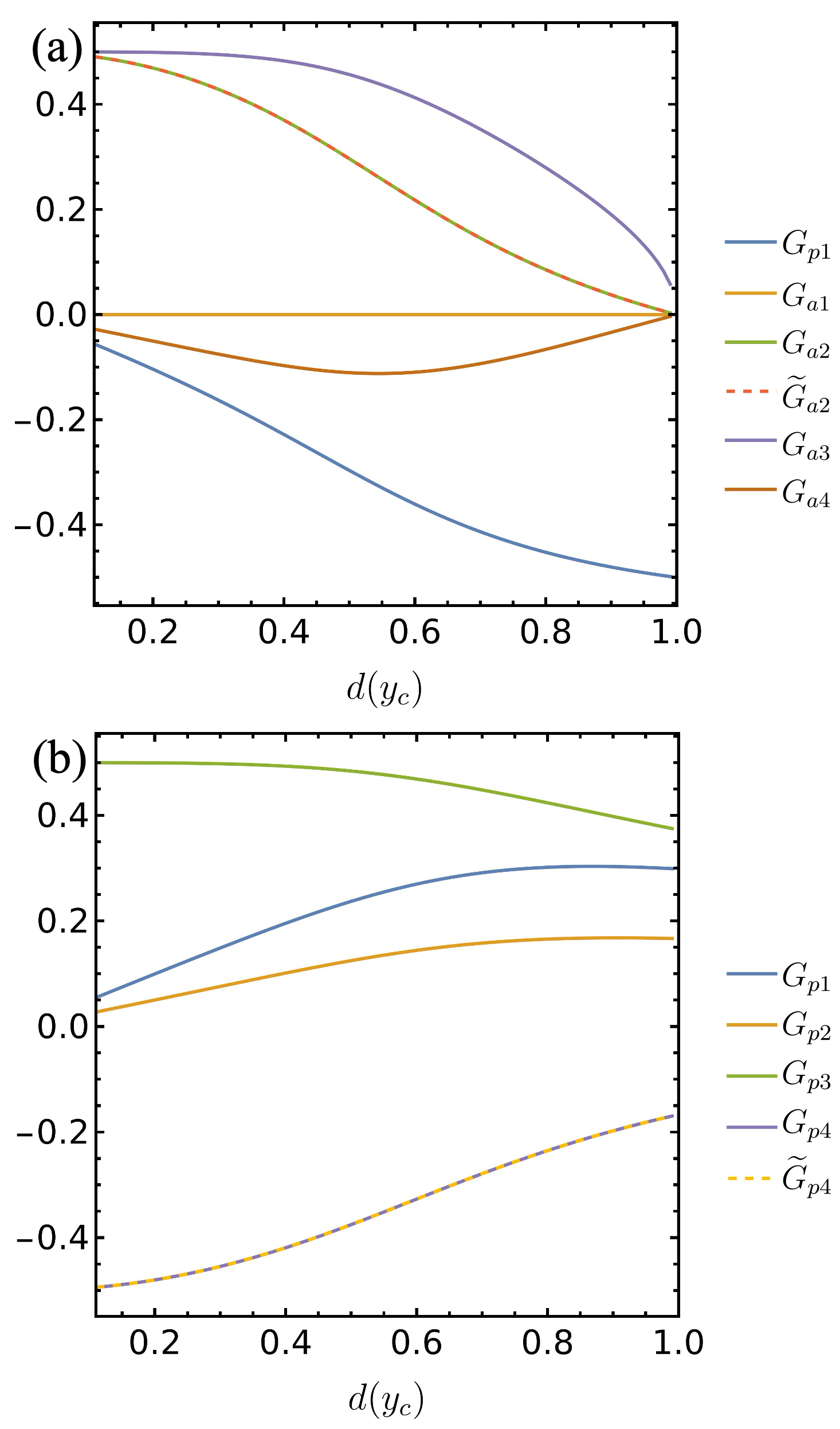}%
    \caption{Coefficients $G_i$ in Eq.~\eqref{eq:coupling-constants-criticaltiy} for the $d_{xy}$-wave AM as a function of $d(y_c)$. (a) the anti-parallel spin solution. (b) parallel spin solutions.}
    \label{fig:coupling-coefficients-dxy}
\end{figure}

Generically, the solutions $g_i(y)$ of the pRG equations diverge at a critical value $y_c$ while their ratios tend to constants upon approaching $y_c$, because of the quadratic form of the differential equations. For $y\rightarrow y_c$, these so-called fixed trajectories are of the form 
\begin{equation}\label{eq:coupling-constants-criticaltiy}
    g_i \approx \frac{G_i}{y_c-y}.
\end{equation}
The values of $G_i$ can be found for fixed $d(y_c)$ by substituting \eqref{eq:coupling-constants-criticaltiy} into the RG equations \eqref{eq:RG-dxy} then solving the coupled algebraic equations for $G_i$ for given $d(y_c)$. For reference we plot $G_i$ as a function of $d(y_c)$ for the two stable fixed-points found in Fig.~\ref{fig:coupling-coefficients-dxy}.

The divergence indicates an instability of the AM metal and we analyze susceptibilities of different orders below to identify the type of instability. For generic initial conditions, there is only a small number of diverging solutions the system flows to. These correspond to stable fixed trajectories which are approached even when perturbing the initial couplings away from them within a given basin of attraction. To find them we start by noting that the coupled equations are invariant under $g_{p3},g_{a3} \rightarrow -g_{p3}, -g_{a3}$. This invariance is lost if there are $n>2$ patches~\cite{Chubukov2012}. This means that RG solutions also appear in pairs which differ by the signs of $g_{p3},g_{a3}$. The $\beta$ functions $\beta_{p3}, \beta_{a3}$ then vanish at $g_{p3} = g_{a3} = 0$. So, the sign of $g_{p3},g_{a3}$ cannot change. Therefore, we always pick the solution with positive $g_{p3}$ or $g_{a3}$ corresponding to repulsive interactions.


We verify in the following that these numerical solutions correspond to the only stable fixed trajectories in the system. For this purpose, we select $g_{p1}$ as the new RG scale $u$ since it diverges in both solutions, and then define the new variables
\begin{equation}\label{eq:RG-variables}
    g_{p1} = u, \ \gamma_i = \frac{g_i}{u};
\end{equation}
here $i=p2,...,a4$. The new variables $\gamma_i$ flow to fixed points with finite values on the fixed trajectory with the relation $\gamma_i=G_i/G_{p1}$ and $G_{p1}=[d(y_c)(-1-\gamma_{a_1}^2+\gamma_{p3}^2-\gamma_{a3}^2+2\gamma_{p2})]^{-1}$ at the fixed point. Their $\beta$-functions are expressed in the original $\beta$-functions $\beta_i$ as
\begin{align}
 \frac{\p\gamma_i}{\p \log u} = \beta_{\gamma_i}= \frac{\beta_i}{\beta_u}(\mathbf{\gamma}) - \gamma_i.  
\end{align}
We denote all $\gamma$ variables as a vector $\mathbf{\gamma}$. The first term on the right is a function of $\mathbf{\gamma}$ only, since $\beta_i$ and $\beta_u$ are both homogeneous functions of second order. Finding all fixed points then reduces to finding numerically all solutions to the algebraic equations
\begin{equation}
    \beta_{\gamma_i}(\mathbf{\gamma}_c)= 0, \ d(y) \rightarrow d(y_c).
\end{equation}
Here we replace $d(y)$ by constant $d(y_c)$ at the critical scale $y_c$; $d(y_c)$ depends on the initial conditions and decreases with increasing AM strength $\lambda$. From these solutions we compute the linearized $\beta$ function matrix at each fixed-point $\mathbf{\gamma}_c$, taking $d(y_c)$ as a parameter
\begin{equation}
  M_{ij}=  \frac{\p \beta_{\gamma_i}}{\p \gamma_j}\bigg|_{\mathbf{\gamma}_c}.
\end{equation}
The stable solutions correspond to those with only non-positive eigenvalues of $M_{ij}$. We find indeed only two stable fixed-points corresponding to the numerical RG results above. 
No new solutions appear using other coupling constants as the RG scale $u$.

Finally, the two solutions above do not tend to the SU$(2)$-symmetric $d$-wave SC solution at $d(y_c)=1$~\cite{Furukawa1998}, where the AM term vanishes. We verify that the SU$(2)$-symmetric solution does exist at $d(y_c)=1$ as an unstable fixed point. Thus, the system is unstable with respect to SU$(2)$ symmetry breaking.

\subsubsection{\texorpdfstring{$d_{x^2-y^2}$}{TEXT}-wave model}

We first consider $\lambda \ll t$ for the $d_{x^2-y^2}$-wave AM where each patch includes both spin-up and -down Fermi surfaces. The RG equations are of an analogous form in the couplings as the ones of the $d_{xy}$ case given by Fig.~\eqref{fig:RG-diagrams-dxy}. However, the loop integrals evaluate to different expressions, cf. Eqs.~\eqref{eq:one-loop-d-wave}-\eqref{eq:one-loop-d-wave-1} and Eqs.~\eqref{eq:one-loop-dxy}-\eqref{eq:one-loop-dxy-1}. Taking this into account, we obtain 
\begin{subequations}\label{eq:RG-d-wave-full}
\begin{align}
    &g_{p1}'= -\widetilde{d}\left(g_{a3}^2 - g_{p3}^2 +g_{p1}^2+g_{a1}^2 - 2g_{p1}g_{p2}\right)\\
    &g_{p2}'= \widetilde{d} \left(g_{p2}^2+g_{p3}^2\right) \\
    &g_{p3}'= \widetilde{d}\left(- g_{p4}g_{p3}+4g_{p2}g_{p3}-2 g_{a1}g_{a3}\right)-\widetilde{g}_{p4}g_{p3}\\
    &g_{p4}'= - \widetilde{d} g_{p4}^2 - g_{p3}^2\\
    &\widetilde{g}_{p4}'= -  \widetilde{g}_{p4}^2 - \widetilde{d} g_{p3}^2\\
    &g_{a1}'= -2\widetilde{d} \left( g_{p1}g_{a1}-g_{a1}g_{p2}\right) \\
    &g_{a2}' = \widetilde{d} g_{a2}^2 + dg_{a3}^2 \\
    &\widetilde{g}_{a2}'= d\widetilde{g}_{a2}^2+\widetilde{d}g_{a3}^2\\
    &g_{a3}'=\widetilde{d}\left( -2g_{a3}g_{a4}+g_{a2}g_{a3} -2 g_{p1}g_{a3}+2 g_{p2}g_{a3}\right)\nn \\
    &+d\widetilde{g}_{a2}g_{a3} \\
    &g_{a4}'=-\widetilde{d}\left(g_{a3}^2 +g_{a4}^2 \right).
\end{align}   
\end{subequations}
As in the $d_{xy}$ case, we have introduced the double-logarithmic variable $y=-\Pi_{\text{pp}}$ and the function $d(y) = \mathrm{d}\Pi_\text{ph}/\mathrm{d}y$ describes imperfect nesting due to finite $\lambda$. Additionally, we introduce the function $\widetilde{d}(y)=\mathrm{d}\Pi/\mathrm{d}y$ for all loop diagrams where both logarithms are cut. For all our results, we approximate $\widetilde{d}(y)= [d(y)]^2$ because at small $y$, $\tilde d(y) \rightarrow 1$, and at large $y$, $\widetilde{d}(y) \sim 1/y$ giving for the magnitude of the corresponding diagrams $\log \log y$, which is almost constant. We checked that using, for example, constant $\widetilde{d}< d$ does not qualitatively change the results.


\begin{figure}
    \centering
    \includegraphics[width=0.8\linewidth]{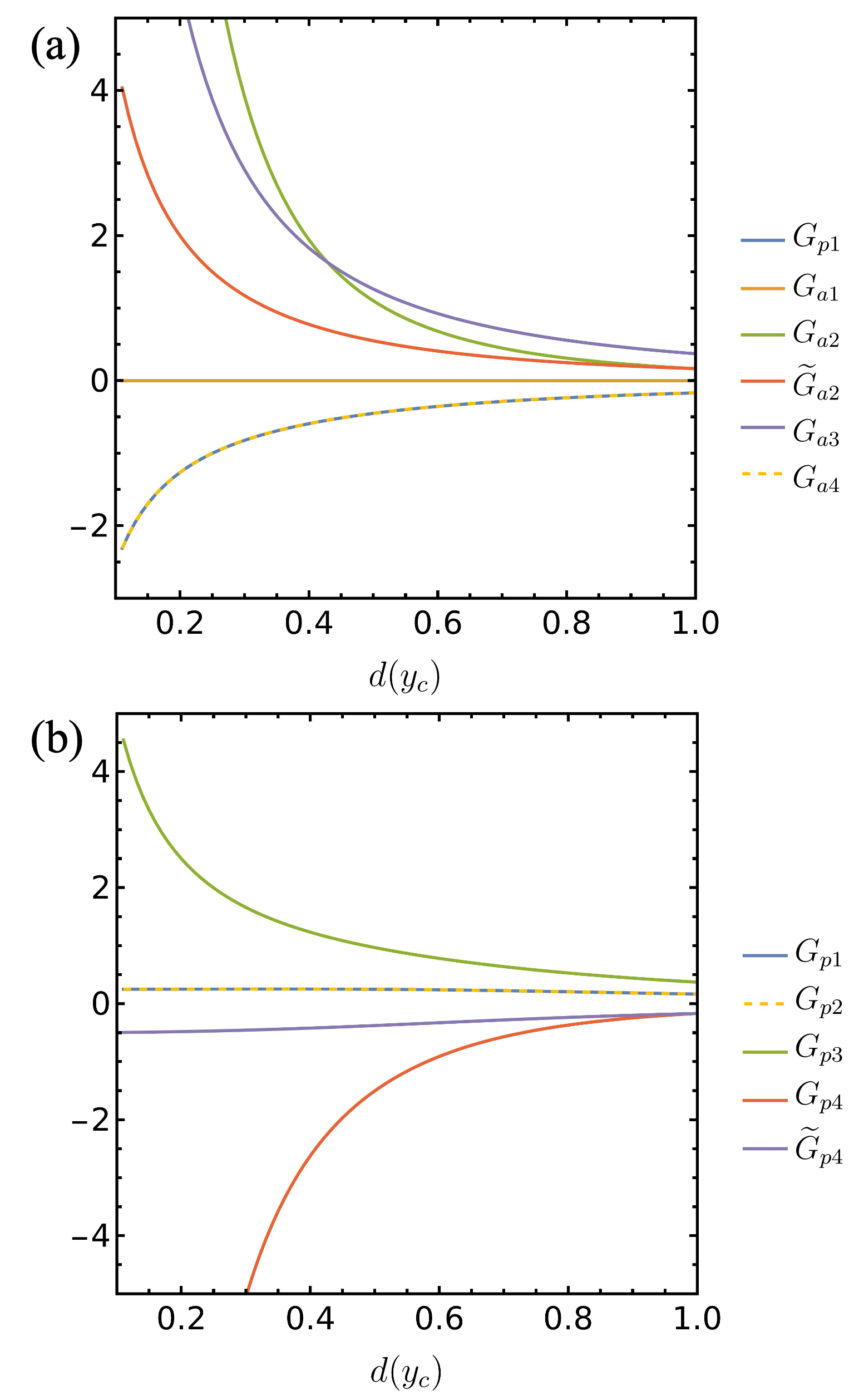}%
    \caption{Coefficients $G_i$ in Eq.~\eqref{eq:coupling-constants-criticaltiy} for the $d_{x^2-y^2}$-wave AM as a function of $d(y_c)$ at $\lambda \ll t$. (a) the anti-parallel spin solution. (b) parallel spin solutions.}
    \label{fig:coupling-coefficients-d-wave}
\end{figure}

In the same way as the previous subsection, we find two stable fixed points in the system with anti-parallel and parallel spin couplings diverging, respectively, see Fig.~\ref{fig:coupling-coefficients-d-wave}. The diverging couplings are  $g_{p1}, g_{a1}, g_{a2}, g_{a3}, \widetilde{g}_{a3}$ for the former, and for the latter $g_{p1}, g_{a1}, g_{a2}, \widetilde{g}_{a2}, g_{a3}$. We also plot $G_i$ as a function of $d(y_c)$ for the two stable fixed points found in Fig.~\ref{fig:coupling-coefficients-d-wave}. These solutions do not tend to the SU$(2)$-symmetric limit, either, i.e., the system is also unstable with respect to SU$(2)$ symmetry breaking. 

For $\lambda \sim t$, we cannot simply take $\widetilde{d}\rightarrow0$ in the RG equations~\eqref{eq:RG-d-wave-full}, since the spin-up Fermi surface at $X$ and spin-down Fermi surface at $Y$ are separated from the VH-points by order of the momentum patch size. This requires considering the two arcs of the Fermi surface near each VH point as separate patches, and the RG equations \eqref{eq:RG-d-wave-full} become invalid. However, with the other Fermi surface far away in energy $2\mu=4\lambda$, only one spin component has its Fermi surface intersect to form a VH point near $X, Y$. Thus, the spin index is also the patch index. This gives rise to a simplified model with only two independent coupling constant $\widetilde{g}_{a2}, \widetilde{g}_{p4}$. Their RG equations are
\begin{equation}\label{eq:RG-d-wave}
   \widetilde{g}_{a2}'= d(y) (\widetilde{g}_{a2})^2; \ \widetilde{g}_{p4}'= -( \widetilde{g}_{p4})^2.
\end{equation}
Thus, due to U(1) spin conversation, the particle-hole and particle-particle channels completely decouple and the RG equations can be integrated directly. Introducing the new variable $\mathrm{d}\xi =d(y)\mathrm{d}y$, the solution is:
\begin{equation}
    \widetilde{g}_{a2} = \frac{ \widetilde{g}_{a2}^{(0)} }{1 -  \widetilde{g}_{a2}^{(0)} \xi}, \  \widetilde{g}_{p4}= \frac{  \widetilde{g}_{p4}^{(0)} }{1 +   \widetilde{g}_{p4}^{(0)}  y},
\end{equation}
where $\widetilde{g}_{a2}^{(0)} $ and $  \widetilde{g}_{p4}^{(0)} $ are bare couplings at the cut-off scale. With repulsive interactions $  \widetilde{g}_{a2}^{(0)} >0$ and $  \widetilde{g}_{p4}^{(0)} >0$ , $\widetilde{g}_{a2}$ diverges whereas $\widetilde{g}_{p4}$ tends to zero under RG flow. This suggests that singlet superconductivity is suppressed because a pairing instability is driven by $\widetilde g_{p4}<0$ and there is no Umklapp coupling $g_{a,p3}$ which could couple particle-hole and particle-particle channel for a Kohn-Luttinger pairing mechanism, see Sec.~\ref{sec:susceptibilities}. Note that, although the RG instabilities are determined by $\widetilde{g}_{a2}, \widetilde{g}_{p4}$, these couplings can in turn drive instabilities of other coupling constants away from the VH points; a similar phenomenon was considered for the hexagonal lattice in Ref.~\onlinecite{Rahul2014}.

\begin{figure}
    \centering
    \includegraphics[width=0.9\linewidth]{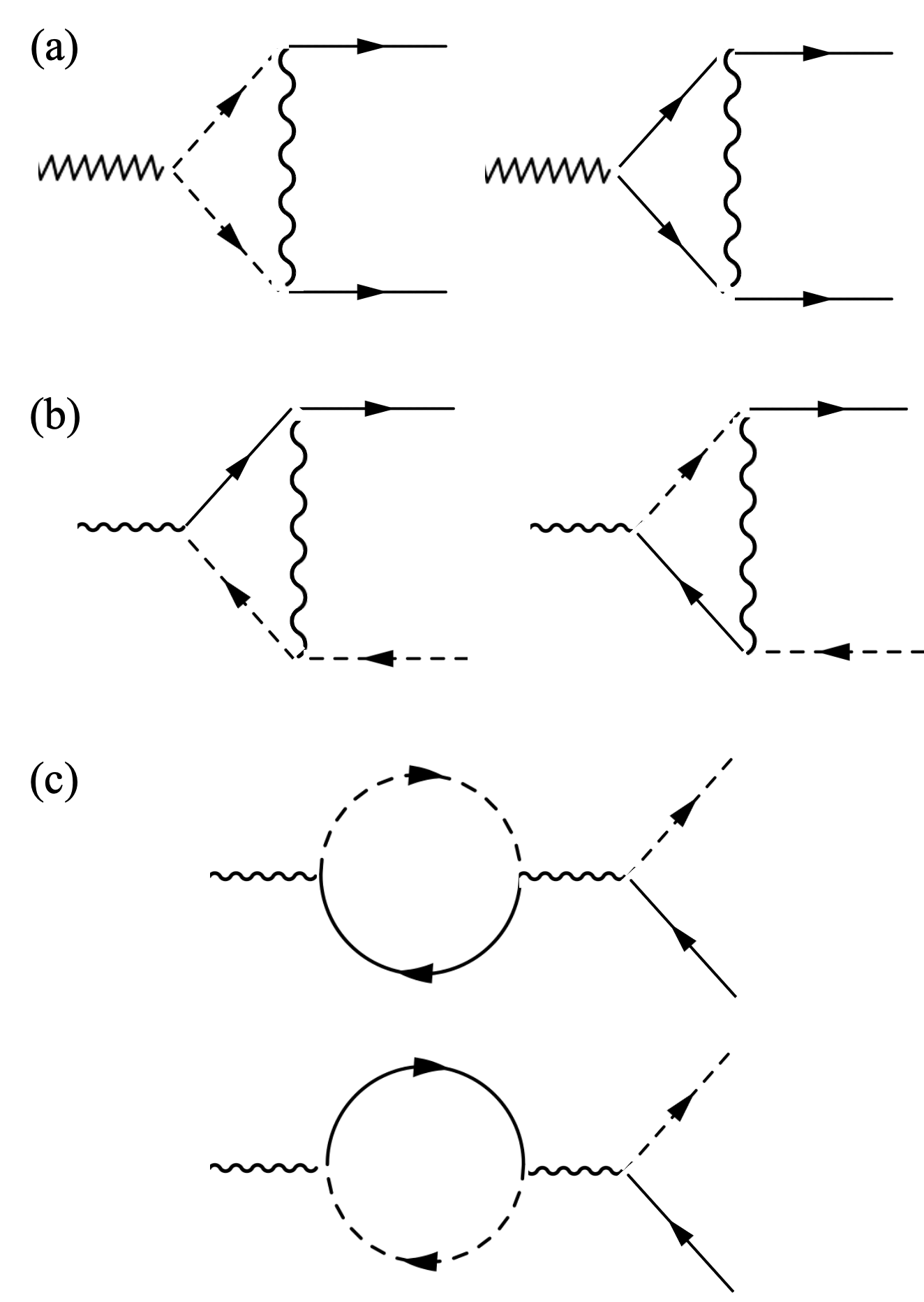}%
    \caption{Feynman diagrams for test vertices. The SC vertex is given by (a). The SDW-t vertex is given by (b). Vertices of CDW and SDW-l receive contributions from both (b) and (c). For SDW-l this arises from $g_{pi}\ne g_{ai}$ due to the breaking of SU$(2)$ symmetry in AM, otherwise the contribution from (c) vanishes. }
    \label{fig:vertex-RG}
\end{figure}

\section{Susceptibilities}\label{sec:susceptibilities}
To test which instabilities the divergence of dressed couplings correspond to, we place the system under weak external perturbations and calculate if the susceptibility for a given ordering tendency also diverges. In the parquet approximation these perturbations appear as vertices $\Gamma$ renormalized by interactions, which diverge in the corresponding instability channel. 
The solution for a given instability has the form~\cite{dzyaloshinskii1972possible,Furukawa1998}:
\begin{equation}\label{eq:vertex-solution}
    \Gamma \sim (y_c-y)^{-\alpha},
\end{equation}
where $\Gamma$ is the corresponding eigenvector and $\alpha$ the eigenvalue. 

Using the exponents $\alpha$ in \eqref{eq:vertex-solution} the corresponding susceptibilities can be computed. If the channel given by vertex $\Gamma$ has double-logarithmic divergence, the corresponding susceptibility is given by
\begin{equation}\label{eq:susceptibility-1}
    \chi = \int_0^{y} \left|\Gamma(y) \right|^2 \mathrm{d}y.
\end{equation}
For channels with imperfect double logarihmic divergences, we have
\begin{equation}\label{eq:susceptibility-2}
    \chi = \int_0^{y} \left|\Gamma(y) \right|^2 d(y) \mathrm{d}y.
\end{equation}
Substituting Eq.~\eqref{eq:vertex-solution}, we find that the divergent solutions at a given critical scale $y_c$ have the form
\begin{equation}\label{eq:susceptibility-solution}
     \chi \sim (y_c-y)^{1-2\alpha},
\end{equation}
where $\alpha>1/2$ is needed for an instability to occur. 

In our AM system the vertices with leading divergences are CDW, transverse spin-density-wave (SDW-t), longitudinal spin-density-wave (SDW-l), and singlet superconductivity (SC); triplet superconductivity cannot be described within the patch model 
because symmetry requires the vertex to vanish on the $X,Y$ points, whereas momentum independent couplings can only give a constant gap near these points.
The considered perturbations are given by adding the following terms to the Hamiltonian
\begin{subequations}\label{eq:vertices-definition}
\begin{align}
   H_{\text{CDW}}&=\Delta_{\text{CDW}}(\mathbf{Q})  \sum_\sigma \psi_{X,\sigma}^\dagger\psi_{Y,\sigma} + \text{h.c.}\\
   H_{\text{SDW-t}}&=m^+ \psi_{X, +}^\dagger \psi_{Y,-} +  m^- \psi_{X,-}^\dagger \psi_{Y,+}+\text{h.c.}\\
   H_{\text{SDW-l}}&=m^z \psi_{X,\alpha}^\dagger \sigma^z_{\alpha \beta} \psi_{Y,\beta} \\
   H_{\text{SC}}&=\Delta_{\text{SC}}(X) \psi_{X,-}^\dagger \psi_{X,+}^\dagger + \Delta_{\text{SC}}(Y) \psi_{Y,-}^\dagger \psi_{Y,+}^\dagger+ \text{h.c.} 
\end{align}
\end{subequations}
with CDW $\Delta_{\text{CDW}}$, transversal magnetic $m^{\pm} = m^x \pm i m^y$, longitudinal magnetic $m^z$, and superconducting $\Delta_{\text{SC}}$ test vertices. Note that the density waves all break tranlsation symmetry, but different additional symmetries: SDW-l does not break any pure spin symmetry, because it is already broken by the AM background. However, it does break the AM $C_4\mathcal T$ symmetry with rotation center on site. SDW-t breaks spin U(1) and AM $C_4\mathcal T$.

We can further distinguish real and imaginary CDW$^{(r,i)}$ and SDW$^{(r,i)}$. In the CDW case, they correspond to the order parameter $\Delta_{\text{CDW}}$ being real or imaginary. Due to the breaking of SU$(2)$ symmetry, SDW-t and SDW-l acquire different exponents. The SDW-l$^{(r,i)}$ correspond to the order
\begin{equation}
    m^{z}\pm m^{z*}.
\end{equation}
The SDW-t pRG solutions are the same with respect to the doublets $(m^{+*}, m^{-})$ and $(m^{+}, m^{-*})$, as a result of the residual U$(1)$ symmetry. We denote SDW-t$^{(r)}$ as the solution which has the eigenvectors $(m^{+}, m^{-*}) =(m^{+*}, m^{-}) = (1,1)/\sqrt{2}$ in the SU$(2)$-limit. They form the U(1) superposition 
\begin{equation}
    c_1 \left( m^{+} + m^{-*} \right) + c_2  \left(  m^{+*} + m^{-} \right),
\end{equation}
and correspond to real transverse SDW as can be seen by setting $c_1=c_2$. The eigenvalue also coincides with that of SDW-l$^{(r)}$ in the  SU$(2)$-limit, signifying real SDW rotated in-plane. Similarly, the imaginary transverse SDW-t$^{(i)}$ is given by the eigenvectors $(m^{+}, m^{-*}) =(m^{+*}, m^{-}) = (1,-1)/\sqrt{2}$ in the SU$(2)$-limit and forms the general superposition
\begin{equation}
    c_1 \left( m^{+} - m^{-*} \right) + c_2  \left(  m^{+*} -  m^{-} \right).
\end{equation}
Lastly, we can distinguish the pairing symmetry by different values of vertices on the patches, which corresponds to the following eigenvectors for $s$- and $d$-wave SC
\begin{equation}
    [ \Delta_{\text{SC}}(X), \Delta_{\text{SC}}(Y)] = \frac{1}{\sqrt{2}}(1,\pm 1).
\end{equation}

In the rest of this section, we present numerical results for the exponent $\alpha$ in Eq.~\eqref{eq:vertex-solution} of all foregoing pRG solutions for the $d_{xy}$-wave and $d_{x^2-y^2}$-wave models. We then discuss the leading instability given by the most strongly divergent susceptibilities.

\begin{figure}
    \centering
    \includegraphics[width=\linewidth]{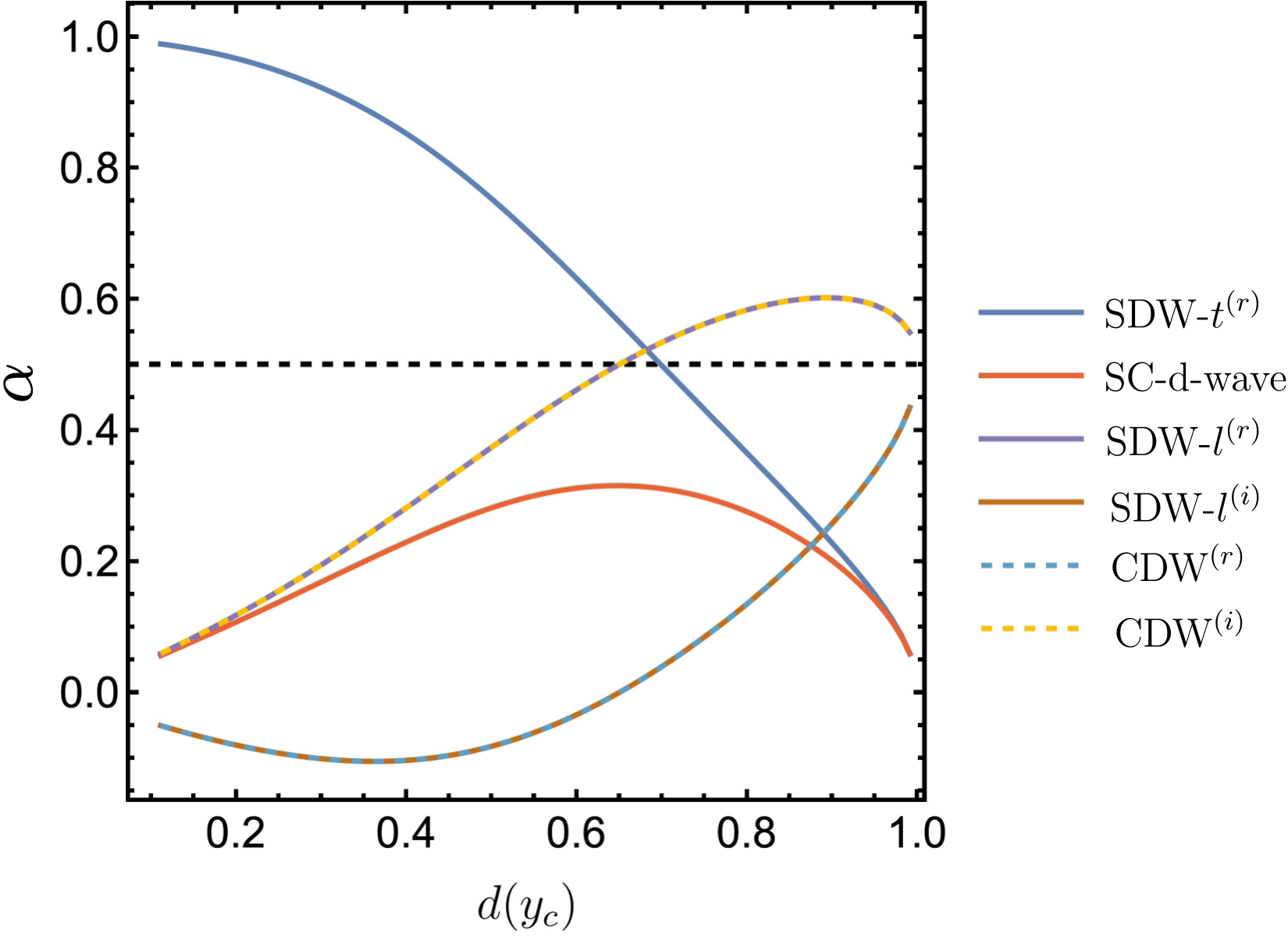}%
    \caption{
    Vertex exponents $\alpha$ of the $d_{xy}$-wave model as a function of $d(y_c)$ for the anti-parallel spin fixed-points; the parallel-spin fixed-points do not give divergent susceptibility. The dashed vertical line corresponds to $\alpha =1/2$ above which the susceptibility diverges in the double-logarithmic approximation. The SDW-l$^{(r)}$ and CDW$^{(i)}$ susceptibilities have the same exponent, so do the SDW-l$^{(i)}$ and CDW$^{(r)}$ susceptibilities. The leading instability at higher $d(y_c)$ should be determined by taking into account the subleading logarithms. Instabilities with $\alpha <0$ for all $d(y_c)$ are not shown.
    }
    \label{fig:vertex-exponents-dxy}
\end{figure}

\subsection{\texorpdfstring{$d_{xy}$}{TEXT}-wave model}
The matrix pRG equations for the vertices in the $d_{xy}$-wave model are diagrammatically depicted in 
Fig.~\ref{fig:vertex-RG}. They are given by
\begin{subequations}\label{eq:vertex-RG-dxy}
\begin{align}
   &\frac{\p }{\p y}\begin{pmatrix}
       \Delta_\text{CDW} \\ \Delta_\text{CDW}^*
   \end{pmatrix}= d(y)\begin{pmatrix}
       g_{p2} - g_{p1}-g_{a1} & -g_{a3} \\-g_{a3} &g_{p2} - g_{p1}-g_{a1}
   \end{pmatrix}
   \begin{pmatrix}
       \Delta_\text{CDW} \\ \Delta_\text{CDW}^*
   \end{pmatrix};\\
   &\frac{\p }{\p y} \begin{pmatrix}
       m^+ \\ m^{-*}
   \end{pmatrix}  =\begin{pmatrix}
       \widetilde{g}_{a2} & g_{a3} \\g_{a3} &g_{a2}
   \end{pmatrix}\begin{pmatrix}
       m^+ \\ m^{-*}
   \end{pmatrix} ;\\
   &\frac{\p }{\p y}\begin{pmatrix}m^z\\m^{z*} \end{pmatrix}
   =d(y)\begin{pmatrix}
       g_{p2} - g_{p1}+g_{a1} & g_{a3} \\g_{a3} &g_{p2} - g_{p1}+g_{a1}
   \end{pmatrix}
   \begin{pmatrix}m^z\\m^{z*} \end{pmatrix};\\
   &\frac{\p }{\p y} \begin{pmatrix}
       \Delta_\text{SC}(X) \\ \Delta_\text{SC}(Y)
   \end{pmatrix}  =- d(y)\begin{pmatrix}
       g_{a4} & g_{a3} \\g_{a3} & g_{a4}
   \end{pmatrix}\begin{pmatrix}
       \Delta_\text{SC}(X) \\ \Delta_\text{SC}(Y)
   \end{pmatrix} .\label{eq:vertex-RG-dxy-SC}
\end{align}
\end{subequations}
Taking complex conjugate for the transverse SDW order parameters gives the same equations but with the doublet $(m^{+*}, m^{-})$. We integrate Eq.~\eqref{eq:vertex-RG-dxy} near the critical scale $y_c$ where we can use the universal form of the fixed trajectory $g_i = {G_i}/({y_c-y})$. The vertex equations have the form 
\begin{equation}\label{eq:vertex-equation}
    \frac{\p \Gamma}{ \p y} = \frac{\alpha}{y_c-y}\Gamma,
\end{equation}
where $\Gamma$ is the eigenvector and $\alpha$ the eigenvalue for the matrices in Eq.~\eqref{eq:vertex-RG-dxy}. Integrating Eq.~\eqref{eq:vertex-equation} gives Eq.~\eqref{eq:vertex-solution}. The exponents for the $d_{xy}$-wave instabilities are
\begin{subequations}\label{eq:vertex-exponents-dxy}
\begin{align}
   &\text{CDW}^{(r)}: \ \alpha = d(y_c)(G_{p2} - G_{p1}-G_{a1}-G_{a3});\\
   &\text{CDW}^{(i)}: \ \alpha = d(y_c)(G_{p2} - G_{p1}-G_{a1}+G_{a3});\\
   &\text{SDW-t}^{(r)}:\  \alpha = \frac{G_{a2} +\widetilde{G}_{a2}}{2}+ \sqrt{ \frac{\left(G_{a2} -\widetilde{G}_{a2}\right)^2}{4}+G_{a3}^2};\\
   &\text{SDW-t}^{(i)}:\  \alpha = \frac{G_{a2} +\widetilde{G}_{a2}}{2}- \sqrt{ \frac{\left(G_{a2} -\widetilde{G}_{a2}\right)^2}{4}+G_{a3}^2};\\
   &\text{SDW-z}^{(r)}: \ \alpha = d(y_c)(G_{p2} -G_{p1}+G_{a1} +G_{a3});\\
   &\text{SDW-z}^{(i)}: \ \alpha = d(y_c)(G_{p2}  -G_{p1}+G_{a1}-G_{a3});\\
   &s\text{-wave SC}: \ \alpha = -\widetilde{d}(G_{a3} +G_{a4});\\
   &d\text{-wave SC}: \ \alpha = d(y_c)(G_{a3} -G_{a4}).
\end{align}
\end{subequations}
In above the superscript $(r,i)$ means the corresponding order-parameter is real or imaginary. As described above the real or imaginary transverse SDW instability corresponds to linear superposition of the eigenvectors for doublets $(m^{+*}, m^{-})$ and $(m^{+}, m^{-*})$ with the same eigenvalues. The degeneracy of eigenvalues for the two doublets is a result of the in-plane $U(1)$ spin rotation symmetry. Due to the breaking of SU$(2)$ symmetry by finite $\lambda$, the longitudinal SDW vertex receives additional contributions from the loop diagrams in Fig.~\ref{fig:vertex-RG}(c) in which, somewhat counterintuitively, anti-parallel spin scattering $g_{a3}$ enhances the instability. In the SU(2)-symmetric limit with $\lambda = 0$, $G_{pi}=G_{ai}$, SDW-l and SDW-t have identical exponents. Thus, the latter are obtained by rotating the real or imaginary longitudinal SDW order in-plane. $d$-wave superconductivity corresponds to the eigenvector $\Delta_{\text{SC}} = (1,-1)$ in Eq.~\eqref{eq:vertex-RG-dxy-SC}. The $s$-wave instability $\Delta_{\text{SC}} = (1,1)$ always has $\alpha<0$ and is thus suppressed.


We plot the exponents $\alpha$ in Eq.~\eqref{eq:vertex-solution} for all leading instabilities in Fig.~\ref{fig:vertex-exponents-dxy} for the anti-parallel spin fixed-points in Fig.~\ref{fig:coupling-coefficients-dxy}(a).
We see from Fig.~\ref{fig:vertex-exponents-dxy} that $\chi^{(r)}_\text{SDW-t}$ becomes the leading instability for a broad range of $d(y_c)\lesssim 0.7$. This is because nesting of anti-parallel spin dispersions remains intact for any AM splitting $\lambda$. For small $\lambda$, i.e., $d(y_c)\rightarrow 1$, equal spin dispersions also become approximately nested again. In this regime for higher $d(y_c)$, we find that $\chi^{(r)}_\text{SDW-l}$ and $\chi^{(i)}_\text{CDW}$ have the most strongly diverging susceptibilities. We note that the degeneracy of SDW-l$^{(r)}$ and CDW$^{(i)}$ will be lifted by including sub-leading contributions.

Interestingly, the parallel fixed points in Fig.~\ref{fig:coupling-coefficients-dxy}(b) do not give rise to divergent susceptibilities within the patch model. However, we cannot rule out other instabilities such as ferromagnetism or triplet pairing, which are subleading or cannot be tested for within a patch approximation; see Appendix~\ref{sec:triplet}. At this fixed point, spin up and down electrons completely decouple, i.e., the system effectively behaves like two copies of spinless fermions. Furthermore, the intra-patch density-density coupling becomes attractive $g_{p4}=\widetilde g_{p4}<0$. This is why we suspect that this fixed point may indicate a triplet pairing instability~\cite{ojajarvi2024pairing}.

\subsection{\texorpdfstring{$d_{x^2-y^2}$}{TEXT}-wave model}

The vertex diagrams for the $d_{x^2-y^2}$-wave model are also given by Fig.~\ref{fig:vertex-RG} taking into account the different expressions for the loop integrals. The corresponding vertex equations for $\lambda \lesssim t$ are:
\begin{subequations}\label{eq:vertex-RG-d-wave-full}
\begin{align}
   &\frac{\p }{\p y}\begin{pmatrix}
       \Delta_\text{CDW} \\ \Delta_\text{CDW}^*
   \end{pmatrix}= \widetilde{d}(y)\begin{pmatrix}
       g_{p2} - g_{p1}-g_{a1} & -g_{a3} \\-g_{a3} &g_{p2} - g_{p1}-g_{a1}
   \end{pmatrix}
   \begin{pmatrix}
       \Delta_\text{CDW} \\ \Delta_\text{CDW}^*
   \end{pmatrix};\\
   &\frac{\p }{\p y} \begin{pmatrix}
       m^+ \\ m^{-*}
   \end{pmatrix}  =\begin{pmatrix}
       d(y)\widetilde{g}_{a2} & \widetilde{d}(y)g_{a3} \\d(y) g_{a3} &\widetilde{d}(y)g_{a2}
   \end{pmatrix}\begin{pmatrix}
       m^+ \\ m^{-*}
   \end{pmatrix} ;\\
   &\frac{\p }{\p y}\begin{pmatrix}m^z\\m^{z*} \end{pmatrix}
   =\widetilde{d}(y)\begin{pmatrix}
       g_{p2} - g_{p1}+g_{a1} & g_{a3} \\g_{a3} &g_{p2} - g_{p1}+g_{a1}
   \end{pmatrix}
   \begin{pmatrix}m^z\\m^{z*} \end{pmatrix};\\
   &\frac{\p }{\p y} \begin{pmatrix}
       \Delta_\text{SC}(X) \\ \Delta_\text{SC}(Y)
   \end{pmatrix}  =- \widetilde{d}(y)\begin{pmatrix}
       g_{a4} & g_{a3} \\g_{a3} & g_{a4}
   \end{pmatrix}\begin{pmatrix}
       \Delta_\text{SC}(X) \\ \Delta_\text{SC}(Y)
   \end{pmatrix} .\label{eq:vertex-RG-d-wave-SC}
\end{align}
\end{subequations}
As a result the exponents for the $d_{x^2-y^2}$-wave instabilities are
\begin{subequations}\label{eq:vertex-exponents-d-wave-full}
\begin{align}
   &\text{CDW}^{(r)}: \ \alpha = \widetilde{d}(G_{p2} - G_{p1}-G_{a1}-G_{a3});\\
   &\text{CDW}^{(i)}: \ \alpha =\widetilde{d}(G_{p2} - G_{p1}-G_{a1}+G_{a3});\\
   &\text{SDW-t}^{(r)}:\  \alpha = \frac{ \widetilde{d} G_{a2} +d \widetilde{G}_{a2}}{2}+ \sqrt{ \frac{\left(\widetilde{d} G_{a2} -d \widetilde{G}_{a2}\right)^2}{4}+d\widetilde{d}G_{a3}^2};\\
   &\text{SDW-t}^{(i)}:\  \alpha = \frac{ \widetilde{d} G_{a2} +d \widetilde{G}_{a2}}{2} - \sqrt{ \frac{\left(\widetilde{d} G_{a2} -d \widetilde{G}_{a2}\right)^2}{4}+d\widetilde{d}G_{a3}^2};\\
   &\text{SDW-l}^{(r)}: \ \alpha = \widetilde{d}(G_{p2} -G_{p1}+G_{a1} +G_{a3});\\
   &\text{SDW-l}^{(i)}: \ \alpha = \widetilde{d}(G_{p2}  -G_{p1}+G_{a1}-G_{a3});\\
   &s\text{-wave SC}: \ \alpha = -\widetilde{d}(G_{a3} +G_{a4});\\
   &d\text{-wave SC}: \ \alpha = \widetilde{d}(G_{a3} -G_{a4}).
\end{align}
\end{subequations}
As before, they reproduce the SU(2) limit with $\lambda = 0$ for $G_{pi}=G_{ai}$, $d=\widetilde{d}=1$, and SDW-l and SDW-t have identical exponents. 

\begin{figure}
    \centering
    \includegraphics[width=0.97\linewidth]{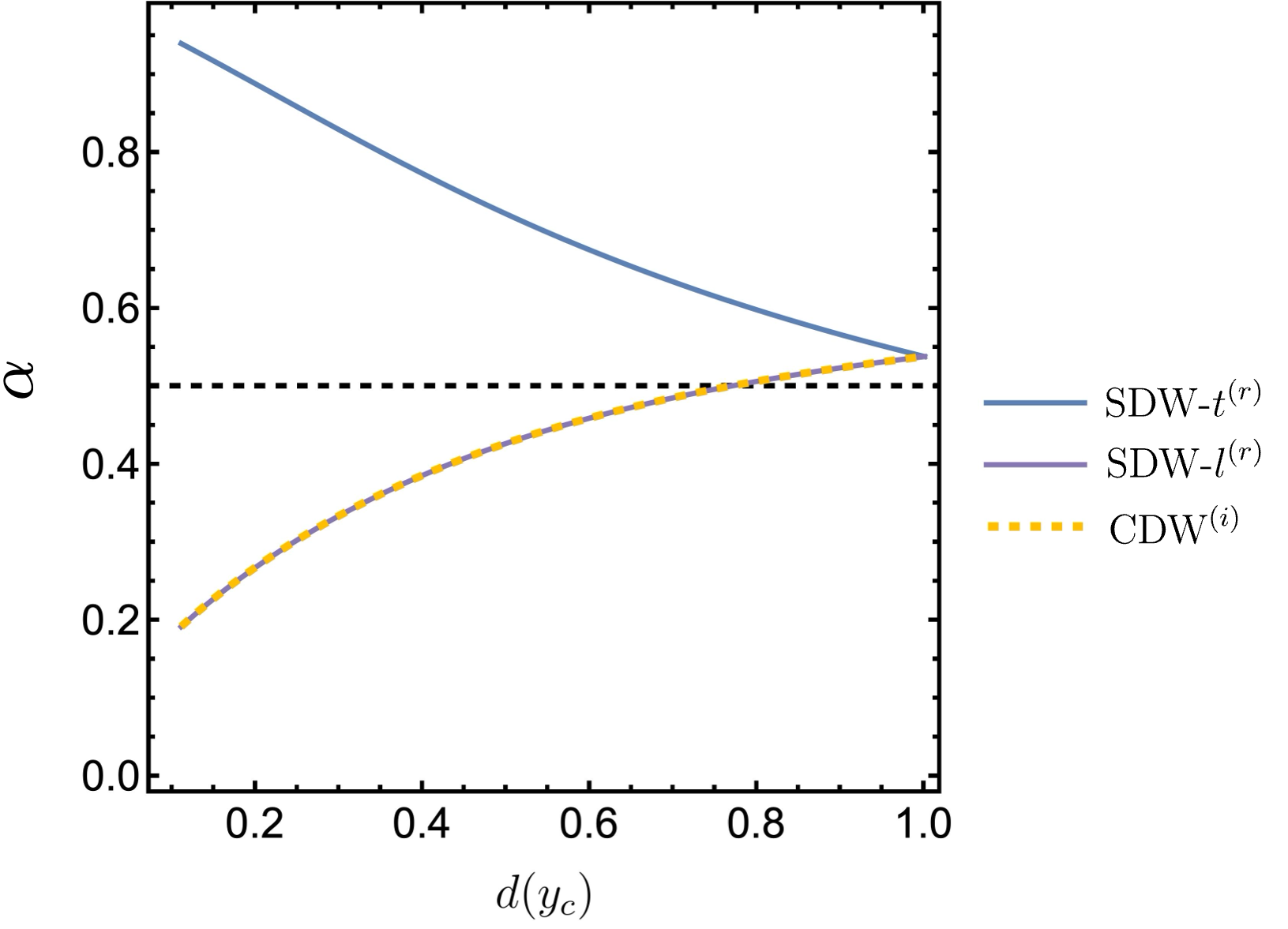}%
    \caption{
    Vertex exponents $\alpha$ of the $d_{x^2-y^2}$-wave model at small $\lambda \ll t$ as a function of $d(y_c)$ for the anti-parallel spin fixed-points; the parallel-spin fixed-points do not give divergent susceptibility. The dashed vertical line corresponds to $\alpha =1/2$ above which the susceptibility diverges in the double-logarithmic approximation. The leading instability is SDW-t$^{(r)}$. Exponents for SDW-l$^{(r)}$, $d$-wave SC and CDW$^{(i)}$ and for $s$-wave SC, SDW-l$^{(i)}$ and CDW$^{(r)}$ coincide. Note that the condition $\lambda \ll t$ implies that our results might not be valid at small $d(y_c)$, but the SDW-t$^{(r)}$ susceptibility is dominant for all $d(y_c)$. Instabilities with $\alpha <0$ for all $d(y_c)$ are not shown.
    }
    \label{fig:vertex-exponents-d-wave}
\end{figure}

For $\lambda\ll t$, we plot the exponent $\alpha$ for all leading instabilities in Fig.~\ref{fig:vertex-exponents-d-wave} for the anti-parallel spin fixed-points in Fig.~\ref{fig:coupling-coefficients-d-wave}(a). We see that the leading instability is always SDW-t$^{(r)}$. This is expected because only the spin-down Fermi surface at $X$ and the spin-up Fermi surface at $Y$ exhibit VH points. Scattering processes of equal spin are logarithmically smaller. As in the $d_{xy}$ case, the parallel fixed point in Fig.~\ref{fig:coupling-coefficients-d-wave}(b) describes a system of decoupled spin up and down fermions with attractive couplings $g_{p4},\widetilde g_{p4}<0$ and we do not find any divergent susceptibilities within the patch model, which leaves triplet pairing as a possible instability.

For $\lambda\sim t$, we consider only $\widetilde{g}_{p4}, \widetilde{g}_{a2}
 \neq0$ as explained before Eq.~\ref{eq:RG-d-wave}. This gives the vertex equations 
\begin{subequations}\label{eq:vertex-RG-d-wave}
\begin{align}
   &\frac{\p m^+}{\p y}=d(y)\widetilde{g}_{a2}m^+ ;\\
   &\frac{\p \Delta_\text{SC}}{\p y} =-\widetilde{g}_{p4}\Delta_\text{SC}.
\end{align}
\end{subequations}
In terms of the coefficients near $y_c$, $G_i = g_i (y_c-y)$, the exponents $\alpha$ can be expressed as 
\begin{subequations}\label{eq:vertex-exponents-d-wave}
\begin{align}
   &\text{SDW-t}:\  \alpha = d(y_c) \widetilde{G}_{a2};\\
   &\text{SC}: \ \alpha = -\widetilde{G}_{p4}.
\end{align}
\end{subequations}
For an attractive interaction $\widetilde g_{p4}<0$ a pairing instability can arise. With repulsive interactions $\widetilde{G}_{a2},\widetilde{G}_{p4}>0$, superconductivity is suppressed and the transverse SDW is again the leading instability.

For the leading instability SDW-t$^{(r)}$, the results can be obtained analytically without needing the approximation \eqref{eq:coupling-constants-criticaltiy} by changing into the new logarithmic variable $\mathrm{d}\xi = d(y)\mathrm{d}y$. From Eq.~\eqref{eq:RG-d-wave}, the coupling constant $\widetilde{g}_{a2}$ has the form
\begin{equation}
    \widetilde{g}_{a2} = \frac{1}{\xi_c-\xi}.
\end{equation}
Eqs.~\eqref{eq:susceptibility-2} and \ref{eq:vertex-RG-d-wave} then give:
\begin{equation}
    m^+ = \frac{1}{\xi_c-\xi}, \ \chi_\text{SDW-t}^{(r)} \sim \frac{1}{\xi_c-\xi}.
\end{equation}
The SC susceptibility is always suppressed for $\widetilde{g}_{p4}>0$. Thus for the whole range of $\lambda$ the leading instability is SDW-t$^{(r)}$.

\section{Discussions}

\subsection{Summary of PRG analysis}

We studied VH-induced instabilities for both $d_{x^2-y^2}$-wave and $d_{xy}$-wave AM metals on a square lattice with $C_4\mathcal{T}$ symmetry. They were proposed in candidate materials such as the V$_2$X$_2$O (X = Te, Se) family~\cite{Smejkal2022Sep,ma2021multifunctional,jiang2025metallic,parthenios2025}, or CoS$_2$ \cite{parthenios2025}. Below we also present a possible scenario based on bilayer cuprates in which the model can be realized.

In the $d_{xy}$-wave AM metal, we find that at half-filling, VH points exist at $X$ and $Y$ points of the BZ. Each VH point has Fermi surfaces of both spin components, rotated by opposite angles determined by the strength of the AM splitting $\lambda$. By solving the resulting pRG equations, we find two stable fixed points. The solution, where predominantly anti-parallel spin couplings diverge, corresponds to an instability towards a real transverse SDW for a broad range of $\lambda$. For small $\lambda$, susceptibilities for real longitudinal SDW and imaginary CDW diverge most strongly. This `degeneracy' can be removed by subleading logarithmic contributions. A SC instability remains absent for bare, repulsive interactions. At the second fixed-point solution, where only parallel spin couplings diverge, we do not find any diverging susceptibility within the patch model. However,  attractive intra-patch density-density couplings may indicate a triplet SC instability.

The $C_4\mathcal{T}$ symmetry also allows for a $d_{x^2-y^2}$-wave AM model. In the $d_{x^2-y^2}$-wave model, the Fermi surface of one spin component loses the VH singularity due to finite $\lambda$. However, at $\lambda \ll t$ the spin-symmetric hopping, the difference between both spin components can be neglected compared to the RG scale (e.g., temperature, external frequency) and the leading instability is still a real transverse SDW. As $\lambda$ increases to order $t$, each VH point has a Fermi surface of only one spin component. The particle-particle channel then completely decouples from the particle-hole channels, since two electrons with parallel spins near one VH cannot be scattered to the other VH point with both spins flipped. In this case also, the leading instability from the anti-parallel-spin fixed point is a real transverse SDW. As in the $d_{xy}$ case, no susceptibility diverges within the patch model, but attractive intra-patch density-density couplings may indicate triplet pairing.

In both models, the leading instabilities are spin or charge density waves. However, as one varies $\mu$ away from VH fillings pairing instabilities can become dominant. This is because $\mu$ serves as an IR regulator of the one-loop diagrams and increasing $\mu$ effectively reduces the doubly-logarithmically divergent diagrams until they become comparable with single logarithmic ones which we have neglected in our analysis. We expect that singlet pairing is still suppressed because of the spin splitting which prevents momentum-reversed states with opposite spins at the Fermi level. Triplet pairing remains a possibility. As discussed in the Introduction and App.~\ref{sec:triplet}, a triplet pairing instability is absent without taking into account momentum-dependent interaction vertices. However, in Ref.~\cite{parthenios2025} it was shown using functional RG (which accounts for the momentum dependence) that for small altermagnetic splitting, neither singlet nor triplet pairing instabilities occur above a temperature of $10^{-3}t$ when tuning away from VH filling. 
In contrast, for large altermagnetic splitting, a pair density wave (PDW) instability in the triplet channel occurs for the $d_{x^2-y^2}$-model away from VH filling.  
Another way to obtain a pairing instability is to introduce a small second nearest-neighbor hopping $t'$ which destroys Fermi surface nesting. In the $d_{xy}$-model fRG also shows that $d$-wave pairing occurs when the nearest-neighbor hopping trumps the effect of the altermagnetic splitting $t'\gtrsim \lambda$~\cite{parthenios2025}. 

\subsection{Realization of the \texorpdfstring{$d_{x^2-y^2}$}{TEXT}-wave AM in a square-lattice model 
}\label{sec:exp-scenario}

\begin{figure}
    \centering
    \includegraphics[width=0.6\linewidth]{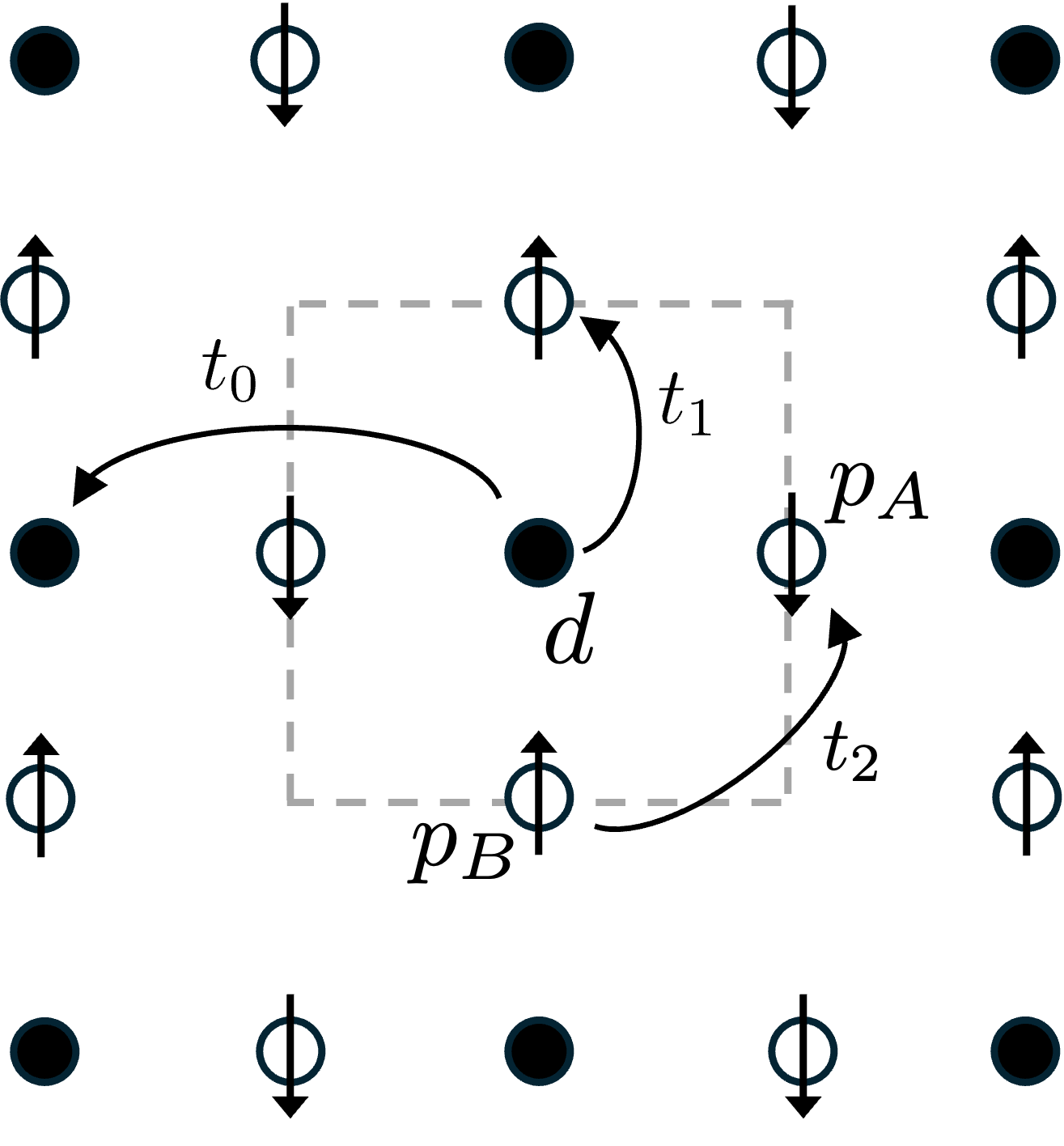}%
    \caption{Lattice model of the Emery model described in Ref.~\onlinecite{Fischer2011} and Sec.~\ref{sec:exp-scenario}. The Cu and O atoms are shown black and white circles respectively corresponding to $d$- and $p$-orbitals. The unit cell is shown with gray dashed lines. The Neel vectors are shown as black arrows. Cu-Cu, Cu-O and O-O hopping are denoted by $t_0, t_1$ and $t_2$ respectively. The secondary instabilities we study lead to additional orders on the Cu lattice. Note that the AM symmetry $C_4\mathcal T$ we defined before would correspond to a rotation center on the Cu atoms.}
    \label{fig:cuprates}
\end{figure}

Possible experimental scenarios for spin-split Fermi surface are proposed on the CuO$_2$ Emery model~\cite{Fischer2011,li2024d}, or more generally on the Lieb lattice~\cite{Brekke2023Dec,Durrnagel2024Dec,kaushal2024altermagnetism}; see Fig.~$2$ and Sec.~III A2 in Ref.~\onlinecite{Fischer2011}. We now show that the `spin-nematic phase' for the Emery model, proposed in Ref.~\onlinecite{Fischer2011}, realizes the $d_{x^2-y^2}$-wave AM and the form factor is given by Eq.~\eqref{eq:form-factor-d-wave}. The CuO$_2$ lattice is shown in Fig.~\ref{fig:cuprates}. The AM phase corresponds to non-zero AFM N\'eel vectors $\mathbf{m}= m\mathbf{e}_z$ on the insulating oxygen atoms. Thus, to a first approximation the copper hopping amplitude receives corrections from virtual transitions to the AFM oxygen. In the Hamiltonian the kinetic term $H_0$ includes nearest-neighbor (NN) hopping between Cu atoms $t_0$, NN Cu-O hopping amplitude $t_1$ and NN O-O amplitude $t_2$:
\begin{equation}
    H_0 = -t_0 \sum_{i,j}d_{\sigma,i}^\dagger d_{\sigma,j}  - t_1 \sum_{i,j} p_{\sigma,i}^\dagger d_{\sigma,j}- t_2 \sum_{i,j} p_{\sigma,i}^\dagger d_{\sigma,j} + \text{h.c.}
\end{equation}
The spin-dependent MF Hamiltonian given by Eq.~(6) in Ref.~\onlinecite{Fischer2011} becomes:
\begin{subequations}
\begin{align}    
&H = \sum_{\mathbf{p},\sigma} \Psi_\sigma^\dagger(\mathbf{p}) \left[H^\sigma_\text{MF}(\mathbf{p})-\mu \right] \Psi_\sigma(\mathbf{p}),\\
&H^\sigma_{\text{MF}}(\mathbf{p})=
    \begin{pmatrix}
        - m U \sigma & \gamma_2(\mathbf{p})& \gamma_1(p_x) \\
       \gamma_2(\mathbf{p})& mU  \sigma &  \gamma_1(p_y)\\
      \gamma_1(p_x)& \gamma_1(p_y) & \gamma_0(\mathbf{p}) \\
    \end{pmatrix} \label{eq:MF-Hamiltonian} \\
&\Psi_\sigma(\mathbf{p}) = [p_{A\sigma}(\mathbf{p}),p_{B\sigma}(\mathbf{p}), d_{\sigma}(\mathbf{p})]^{\text{T}}.
\end{align}    
\end{subequations}
Here $U$ is given by the interaction energy scale associated with the AFM order and Cu-Cu and Cu-O hopping terms:
\begin{equation}\label{eq:hopping-term}
\begin{split}
    &\gamma_0(\mathbf{p}) = -2t_0 (\cos p_x + \cos p_y), \ \gamma_1(p) = -2t_1 \cos \frac{p}{2},\\
    &\gamma_2(\mathbf{p}) = -2t_2 \left(\cos \frac{p_x+p_y}{2}+\cos \frac{p_x-p_y}{2}\right).
\end{split}
\end{equation}
We shall assume that the oxygen atoms are strongly insulating, i.e. their band gap $mU$ satisfies $mU \gg t_0, t_1, t_2$. In this limit we can neglect $\gamma_2$ and the Cu band is modified due to virtual transitions to the oxygen atoms. The MF Hamiltonian \eqref{eq:MF-Hamiltonian} then consists of two $p$-orbital sectors at energy $\pm mU$ for each spin component, which are coupled perturbatively to the $d$-orbital sector via $\gamma_1$. This allows one to project out the $p$-orbital sectors by using the effective Hamiltonian formula 
\begin{equation}
    (H_\text{eff})_{mn} =(H_0)_{mn} + \frac{1}{2}\sum_{l} V_{nl}V_{lm} \left( \frac{1}{E_n-E_l} + \frac{1}{E_m-E_l}\right),
\end{equation}
where $H_0$ is the Hamiltonian of the low-energy sector of interest (here the $d$ electrons with bandwidth $\sim t_0$) with states $n,m$ and energies $E_n, E_m$. The summation is over eigenstates $l$ of other sectors ($p$ electrons) with the corresponding energies $E_l=\pm mU$. $V\sim t_1$ is the perturbation such that $|V_{nl}/(E_l-E_n)|\ll 1$. For Eq.~\eqref{eq:MF-Hamiltonian} this gives the effective Hamiltonian for the $d$-orbitals to second order in $t_1/U$:
\begin{equation}
    H_\text{eff}^{\sigma} = \gamma_0(\mathbf{p})  - \frac{1}{ mU} \sigma \left( |\gamma_+(\mathbf{p})|^2 -  |\gamma_-(\mathbf{p})|^2 \right),
\end{equation}
Substituting Eq.~\eqref{eq:hopping-term} gives:
\begin{equation}
 H_\text{eff}^{\sigma} = \gamma_0(\mathbf{p})+ \frac{4t_1^2}{mU} \sigma^z (\cos p_x -\cos p_y).
\end{equation}
The spin-dependent term has the form of \eqref{eq:form-factor-d-wave}.

\subsection{Conclusions}

We studied a VH scenario for 2D altermagnetic metals with $C_4\mathcal{T}$ symmetry and $d_{xy}$ or $d_{x^2-y^2}$ AM splitting via PRG for the corresponding patch models. For both models, we find secondary SDW as the leading instability. Our predictions of leading SDW and absence of pairing instabilities at VH filling agree with Ref.~\onlinecite{parthenios2025} which considers the same system using the FRG method. 

Furthermore, we obtain a solution with divergent parallel-spin couplings that remains in a critical state without singular susceptibilities for spin, charge, or superconducting orders within a patch approximation. We argued that it may enhance triplet SC which cannot be described with the approach of this paper and is a subject for future work. We also note that a system with parallel-spin couplings only is formally identical to two copies of spin-less fermions.

Finally, we recover the SU$(2)$-invariant solution for both models as an unstable fixed-point solution at $\lambda = 0$. This suggests that the square lattice VH system is unstable with respect to spin SU$(2)$ symmetry-breaking by AM. In particular, $d$-wave superconductivity never becomes dominant for any $0<d(y_c)<1$, whereas on the square lattice it is the dominant instability in the spin-degenerate limit.

Our results apply to 2D altermagnetic metals with high transition temperatures. This includes the material family of V$_2$X$_2$O (X=Se,Te), or CoS$_2$. Interestingly, a secondary SDW transition was recently reported in KV$_2$Se$_2$O~\cite{jiang2025metallic}. We also proposed a realisation of our model in a designer $d_{xy}$ AM based on two square-lattice layers, one with antiferromagnetic spin arrangement and one in a metallic state. In the future, it will be interesting to include sublattice or orbital degrees of freedom in an effective, minimal description for the VH scenario in AM metals and to extend our approach to other symmetries or lattices.

\begin{acknowledgments}
We thank A.V. Chubukov, Nikolaos Parthenios, and J{\"o}rg Schmalian for useful discussions. J.K. acknowledges support from the Deutsche Forschungsgemeinschaft (DFG, German Research Foundation) under Germany’s Excellence Strategy (EXC–2111–390814868 and ct.qmat EXC-2147-390858490), and DFG Grants No. KN1254/1-2, KN1254/2-1 TRR 360 - 492547816 and SFB 1143 (project-id 247310070). J.K. further acknowledges support from the Imperial-TUM flagship partnership. The research is part of the Munich Quantum Valley, which is supported by the Bavarian state government with funds from the Hightech Agenda Bayern Plus. LC was supported by a grant from the Simons Foundation SFI-MPS-NFS-00006741-11.
\end{acknowledgments}

\section*{Data availability}
The Mathematical notebook that numerically solves the RG equations and produces the plots in this article is available at \cite{data} upon reasonable request.

\appendix

\section{Evaluation of one-loop diagrams}\label{sec:one-loop}
We first derive the particle-particle diagram in Eq.~\eqref{eq:one-loop-dxy-BCS}, which is given by
\begin{equation}
    \Pi^{\text{(p)}}_{\text{pp}}(\omega)=-\frac{T}{4\pi^2}\sum_{p_0} \int \frac{\mathrm{d}^2p}{(ip_0-\varepsilon_{X,+})[i(p_0+\omega)+\varepsilon_{X,+}]},
\end{equation}
where $p_0 = (2n+1) \pi T$ is the fermion Matsubara frequency, and $\varepsilon_{X,+}$ is given by Eq.~\eqref{eq:nesting-dxy-1}. Summing over $p_0$ and performing the analytical continuation $i\omega \rightarrow \omega$ gives
\begin{equation}
    \Pi^{\text{(p)}}_{\text{pp}}(\omega)=\frac{1}{4\pi^2}\int \frac{\tanh\left(\varepsilon_{X,+}/2T\right)}{2t(p_y^2-p_x^2) + 2\lambda p_xp_y+\omega}\mathrm{d}^2p.
\end{equation}
By a rotation the quadratic form in momenta in the denominator of the integrand can be brought into diagonal form: $\sqrt{t^2+(\lambda/2)^2}(p_x'^2 - p_y'^2)$. Defining new variables: $\xi_{\pm} = [t^2+(\lambda/2)^2]^{1/4} (p_x'\pm p_y')$, we get
\begin{equation}
    \Pi^{\text{(p)}}_{\text{pp}}(\omega)=\frac{1}{8\pi^2\sqrt{t^2+(\lambda/2)^2}}\int \frac{\tanh\left(\frac{\xi_+\xi_-}{2T}\right)}{2\xi_+\xi_- +\omega}\mathrm{d}\xi_+\mathrm{d}\xi_-.
\end{equation}
The denominator is thus regularized by $\max \{\omega,T \}$. We then make use of a further substitution $\varepsilon= \xi_+\xi_-$ to get approximately
\begin{equation}
    \Pi^{\text{(p)}}_{\text{pp}}(\omega)\approx \frac{1}{4\pi^2\sqrt{t^2+(\lambda/2)^2}}\int^{\sqrt{\Lambda}}_{\max \{\omega,T \}/\sqrt{\Lambda}} \frac{\mathrm{d}\xi_+}{\xi_+}\int^{\sqrt{\Lambda}\xi_+}_{\omega} \frac{\mathrm{d}\varepsilon}{\varepsilon}.
\end{equation}
Performing the integration gives Eq.~\eqref{eq:one-loop-dxy-BCS}. We see that one logarithm comes from approaching the Fermi surface $\varepsilon=0$ in the $\varepsilon$ integration, while another logarithm is due to the VH-point at the origin at $\varepsilon=0$. Up to a minus sign, the nesting diagram $\Pi^{\text{(ap)}}_{\text{ph}}$ in Eq.~\eqref{eq:one-loop-dxy-nesting} is given by the same integral. Expressions at the other VH point are the same due to $C_4\mathcal{T}$ symmetry. 

For the particle-particle diagram in Eq.~\eqref{eq:one-loop-dxy-1-BCS} we have
\begin{equation}\label{eq:one-loop-dxy-2}
\begin{split}
   & \Pi^{\text{(ap)}}_{\text{pp}}(\omega)=-\frac{T}{4\pi^2}\sum_{p_0} \int \frac{\mathrm{d}^2p}{(ip_0-\varepsilon_{X,+})[i(p_0+\omega)+\varepsilon_{X,-}]}\\
    & = \frac{1}{8\pi^2}\int\left[\tanh\left(\frac{\varepsilon_{X,+}}{2T}\right)+\tanh\left(\frac{\varepsilon_{X,-}}{2T}\right) \right] \frac{\mathrm{d}^2p}{2t(p_y^2-p_x^2)+\omega}.
    \end{split}
\end{equation}
The infra-red regulator is given by $\max \{\lambda, \omega,T \}$ since $\lambda$ cuts the double logarithm. Then the region of integration is determined by $\varepsilon_{X,+}>0, \varepsilon_{X,-}>0$ and $\varepsilon_{X,+}<0, \varepsilon_{X,-}<0$. 

To evaluate the integral, we take the small $\lambda$ limit in what follows. The integration limit $\varepsilon_{X,+}>0, \varepsilon_{X,-}>0$ can be approximated as $(t+\lambda/2)p_y^2 > (t-\lambda/2)p_x^2$. Similarly $\varepsilon_{X,+}<0, \varepsilon_{X,-}<0$ becomes $(t-\lambda/2)p_y^2 < (t+\lambda/2)p_x^2$. From Fig.~\ref{fig:Fermi-surface-dxy}(b) it is clear that the integral is formally identical to the spin-degenerate square-lattice particle-hole diagram with imperfect nesting due to finite $\lambda/2$~\cite{Furukawa1998}. More precisely, the limit $(t-\lambda/2)p_y^2 < (t+\lambda/2)p_x^2$ becomes $p_x > (1 - \lambda') |p_y|$ and $p_x < -(1 -\lambda') |p_y| $ where $\lambda' = \lambda/2t$. For $\omega < \lambda$, direct integration gives:
\begin{equation}
\begin{split}
&\frac{1}{8\pi^2 t} \int \mathrm{d} p_y \left(\int_{(1-\lambda')|p_y|}^{\sqrt{\Lambda/t}} +\int^{-(1-\lambda')|p_y|}_{-\sqrt{\Lambda/t}} \right)\frac{\mathrm{d} p_x}{p_y^2-p_x^2} \\
=&-\frac{1}{8\pi^2 t} \log \left(\frac{\lambda}{t}\right)\int \frac{\mathrm{d} p_y}{|p_y|}.
\end{split}
\end{equation}
For the other limit the integral is the same with $p_x$ and $p_y$ exchanged. Integrating them with the low energy cut-off $\max \{\omega,T\}/\sqrt{\Lambda/t}$ immediately gives Eq.~\eqref{eq:one-loop-dxy-1-BCS}. Here, the logarithm coming from the pole in Eq.~\eqref{eq:one-loop-dxy-2} is lost because areas of integration are separated from the integrand pole $p_y^2=p_x^2$ due to the AM splitting. The logarithm from the VH-singularity is still present. 

The particle-particle diagram $\Pi_{\text{pp}}$ in Eq.~\eqref{eq:one-loop-d-wave-BCS} can be calculated in the same way as $\Pi^{\text{(p)}}_{\text{pp}}(\omega)$. For the particle-hole diagram~\eqref{eq:one-loop-d-wave-nesting}, repeating the steps as in the $d_{xy}$-wave model above we obtain
\begin{equation}\label{eq:one-loop-d-wave-1}
\begin{split}
   \Pi^{\text{(ap)}}_{\text{pp}}(\omega)=&- \frac{1}{16\pi^2}\int\left[\tanh\left(\frac{\varepsilon_{Y,+}}{2T}\right)+\tanh\left(\frac{\varepsilon_{X,-}}{2T}\right) \right] \\
   &\times \frac{\mathrm{d}^2p}{2t(p_x^2-p_y^2)+\omega}.
    \end{split}
\end{equation}
The integration is restricted to  $\varepsilon_{X,-}>0, \varepsilon_{Y,+}<0$ and $\varepsilon_{X,-}<0, \varepsilon_{Y,+}>0$. From Fig.~\ref{fig:Fermi-surface-d-wave}(a) these two regions are equivalent to $\varepsilon_{X,-}>0$ and $\varepsilon_{Y,+}>0$. For the first limit we have 
\begin{equation}\label{eq:one-loop-d-wave-2}
\begin{split}
   \Pi^{\text{(ap)}}_{\text{pp}}(\omega)=\frac{1}{8\pi^2}\int \frac{\tanh\left(\varepsilon_{X,-}/2T\right)}{2\varepsilon_{X,-} + \lambda p^2+\omega}\mathrm{d}^2p.
    \end{split}
\end{equation}
This is the same integral as in Eq.~\eqref{eq:one-loop-dxy-2} and Ref.~\onlinecite{Furukawa1998}. The same holds for the limit $\varepsilon_{Y,+}>0$. Together they give the result in Eq.~\eqref{eq:one-loop-d-wave-nesting}.

Other one-loop diagrams that involve states near VH points in the $d_{x^2-y^2}$-wave have both logarithms cut by $\lambda$. We demonstrate this for the particle-particle diagram of opposite spins at patch $X$; the remaining one-loop diagrams behave in an analogous way. The integral is given by 
\begin{align}
  &\Pi=\notag \\
  &\frac{1}{8\pi^2}\int\left[\tanh\left(\frac{\varepsilon_{X,+}}{2T}\right)+\tanh\left(\frac{\varepsilon_{X,-}}{2T}\right) \right]  \frac{1}{2t(p_y^2-p_x^2) - \omega -4\lambda}\mathrm{d}^2p.
\end{align}
The integration region is restricted to $\varepsilon_{X,+}, \varepsilon_{X,-}>0$ and $\varepsilon_{X,+}, \varepsilon_{X,-}<0$. Even though, $ \varepsilon_{X,-}$ contains the VH singularity, the divergence of the denominator $2t(p_y^2-p_x^2)= 4\lambda$ is separated from the VH point by order $\sqrt{\lambda/t}$ in momentum space. The $4\lambda$ factor also cuts off the divergence from approaching the Fermi surface. Thus, we find both logarithms to be cut by the factor $\log^2 (\lambda/t)$, as given in Eq.~\eqref{eq:one-loop-d-wave-subleading}.

\section{Triplet superconducting instability}\label{sec:triplet}
In this appendix we show that triplet superconductivity instability cannot be be studied using the patch model with momentum independent interaction vertices. For this purpose we write $g_{i}(\mathbf{p})$ as functions of momenta.

Let us consider the $X$ patch and both spins up. For spin $S=1$, the gap function satisfies:
\begin{equation}
    \Delta_{X,++} (-\mathbf{k})=-\Delta_{X,++} (\mathbf{k}).
\end{equation}
Due to this symmetry, in the gap equation given by the vertex diagram Fig.~\ref{fig:vertex-RG}(a) the interaction line must be replaced by the anti-symmetrized vertex~\cite{ojajarvi2024pairing}:
\begin{equation}
\begin{split}
     &\Delta_{X,++}(\mathbf{k}) = \frac{T}{2}\sum_{p_0}\int\frac{\mathrm{d}^2p}{(2\pi)^2} \\ 
     &\Big\{\left[g_{p4}(\mathbf{p}+\mathbf{k}) - g_{p4}(\mathbf{p}-\mathbf{k})\right] G_X(p-k)G_X(-p+k)\Delta_{X,++} (\mathbf{p})  \\ 
     &+\left[g_{p3}(\mathbf{p}+\mathbf{k}) - g_{p3}(\mathbf{p}-\mathbf{k})\right] G_Y(p-k)G_Y(-p+k)\Delta_{Y,++} (\mathbf{p})\Big\}.
\end{split}
\end{equation}
where $p_0$ is the loop Matsubara frequency and $G_{X,Y}(p)$ are the Green's functions near patch $X,Y$. For a constant interaction vertex this vanishes identically.


%

\end{document}